\documentclass[sigconf]{acmart}

\renewcommand\footnotetextcopyrightpermission[1]{} 
\setcopyright{none}

\copyrightyear{2021}
\acmYear{2021}

\usepackage{amsmath,amsfonts}
\usepackage{algorithmic}
\usepackage{graphicx}
\usepackage{textcomp}
\usepackage{xcolor}
\usepackage{cramped}
\usepackage{balance} 

\usepackage{amsthm}

\newcommand{\Comment}[1]{\relax}
\newcommand{\ignore}[1]{}

\usepackage{graphicx}
\usepackage{enumitem}
\usepackage{hyperref}
\usepackage{xspace}
\usepackage{color}
\usepackage{framed}
\usepackage[normalem]{ulem}
\usepackage{float}
\usepackage{epstopdf}
\usepackage{caption}
\usepackage{subcaption}
\usepackage{chngpage, lipsum}
\usepackage{stmaryrd}
\usepackage[linesnumbered,lined,ruled, noend]{algorithm2e}
\usepackage{adjustbox}
\usepackage{multirow}
\usepackage{setspace}
\usepackage{pifont}

\makeatletter
\def\@copyrightspace{\relax}
\makeatother

\newcommand\resetnobreak{\@nobreakfalse}

\makeatletter
\newcommand{\removelatexerror}{\let\@latex@error\@gobble}
\makeatother

\newcommand{\Pone}{\ensuremath{\class{P_1}}}
\newcommand{\Ptwo}{\ensuremath{\class{P_2}}}
\newcommand{\Pany}{\ensuremath{\class{P_i}}}

\newcommand{\hcluster}{\ensuremath{\class{PHC}}}
\newcommand{\opt}{\ensuremath{\class{OPT}}}

\newcommand{\wine}{\texttt{Wine}}
\newcommand{\iris}{\texttt{Iris}}
\newcommand{\bcancer}{\texttt{Cancer}}
\newcommand{\heart}{\texttt{Heart}}

\newcommand{\disjoint}{\ensuremath{\mathsf{PCure0}}}
\newcommand{\jointSemi}{\ensuremath{\mathsf{PCure1}}}
\newcommand{\jointFull}{\ensuremath{\mathsf{PCure2}}}
\newcommand{\CURE}{\ensuremath{\mathsf{CURE}}}

\newcommand{\class}[1]{{\ensuremath{\mathsf{#1}}}}
\newcommand{\gen}{\ensuremath{\class{Gen}}\xspace}

\newcommand{\Z}{\mathbb{Z}}

\newcommand{\cc}[1]{\class{#1}}

\newcommand{\dist}{\class{dist}}

\newcommand{\setup}{\class{Setup}}
\newcommand{\cluster}{\class{Cluster}}
\newcommand{\filter}{\class{Output}}

\newcommand{\larr}{\ensuremath{\leftarrow}}

\newcommand{\R}{\ensuremath{\mathcal{R}}}

\newcommand{\dataset}{\ensuremath{\mathcal{D}}}

\newcommand{\vJ}{\ensuremath{\textsc{\bf J}}}
\newcommand{\vE}{\ensuremath{\textsc{\bf E}}}
\newcommand{\vV}{\ensuremath{\textsc{\bf V}}}
\newcommand{\vD}{\ensuremath{\textsc{\bf D}}}

\newcommand{\vL}{\ensuremath{\textsc{\bf L}}}

\newcommand{\vS}{\ensuremath{\textsc{\bf S}}}
\newcommand{\vB}{\ensuremath{\textsc{\bf B}}}
\newcommand{\vR}{\ensuremath{\textsc{\bf R}}}

\newcommand{\vQ}{\ensuremath{\textsc{\bf Q}}}
\newcommand{\vH}{\ensuremath{\textsc{\bf H}}}

\newcommand{\alina}[1]{{\textcolor{red} {Alina says: #1}}}

\newcommand{\sk}{\ensuremath{sk}}
\newcommand{\pk}{\ensuremath{pk}}

\renewcommand{\pk}{\ensuremath{{\sf pk}}}
\renewcommand{\sk}{\ensuremath{{\sf sk}}}

\SetKwInput{Input}{Input}
\SetKwInOut{Output}{Output}
\SetKwInput{PoneInput}{$\Pone$'s Input}
\SetKwInput{PtwoInput}{$\Ptwo$'s Input}
\SetKwInOut{PtwoOutput}{$P_2$'s Output}
\SetKwInOut{PoneOutput}{$P_1$'s Output}
\SetKwInput{COutput}{Output}
\SetKwInput{CPara}{Parameters}
\newcommand{\linespace}{ \vspace{.7em}}
\newcommand{\enOne}[1]{\ensuremath{[#1]}}
\newcommand{\enTwo}[1]{\ensuremath{\llbracket#1 \rrbracket}}
\newcommand{\en}[1]{\ensuremath{[#1]}}
\newcommand{\de}[1]{\ensuremath{\langle#1\rangle}}

\renewcommand{\max}{\cc{max}}
\renewcommand{\min}{\cc{min}}

\newcommand{\minselect}{\cc{MinSelect}}

\newcommand{\maxDist}{\cc{MaxDist}}
\newcommand{\minDist}{\cc{MinDist}}
\newcommand{\argMinSelect}{\cc{ArgMin}}

\newlength{\protowidth} \protowidth \linewidth

\newcommand{\view}{\mathsf{view}}
\newcommand{\adv}{\ensuremath{\mathcal{A}}}
\newcommand{\simu}{\ensuremath{\mathsf{Sim}}}
\newcommand{\secp}{\lambda}

\newcommand{\rowmin}{\mathsf{RowMin}}

\newcommand{\base}{{$\hcluster$}}

\newif\iffull    
\fulltrue
\usepackage{url}

\begin{document}
\fancyhead{}
\def\UrlBreaks{\do\/\do-}

\title{Private Hierarchical Clustering and Efficient Approximation}

\author{Xianrui Meng}
\email{xianru@amazon.com}
\affiliation{%
  \institution{Amazon Web Services}
    \country{}
}
\author{Dimitrios Papadopoulos}
\email{dipapado@cse.ust.hk}
\affiliation{%
  \institution{Hong Kong University of Science and Technology}
    \country{}
}
\author{Alina Oprea}
\email{a.oprea@northeastern.edu}
\affiliation{%
  \institution{Northeastern University}
  \country{}
}
\author{Nikos Triandopoulos}
\email{ntriando@stevens.eduu}
\affiliation{%
  \institution{Stevens Institute of Technology}
    \country{}
}

	\begin{abstract}
In 
\emph{collaborative learning},  multiple parties
contribute their  datasets to jointly deduce global machine learning
models for numerous predictive tasks. Despite its efficacy, this learning paradigm fails to
encompass critical application domains that involve highly sensitive data, such as healthcare and security analytics, where privacy risks limit entities to individually train models using only their own datasets.
In this work, we target
\emph{privacy-preserving collaborative hierarchical clustering}. We introduce a {formal security definition} that aims to achieve balance between utility and privacy
and present a {two-party protocol} that provably satisfies it.
We then extend our protocol with: (i) an {optimized version for single-linkage clustering}, and (ii) {scalable approximation variants}. We implement all our schemes and experimentally evaluate their performance and accuracy on synthetic and real datasets, obtaining very encouraging results. For example, end-to-end execution of our secure approximate protocol for over $1$M $10$-dimensional data samples requires $35$sec of computation and achieves $97.09\%$ accuracy.
\end{abstract}

\ccsdesc[500]{Security and privacy~Cryptography}
\ccsdesc[500]{Computing methodologies~Unsupervised learning}


\keywords{secure computation; private hierarchical clustering; secure approximation
} 

\maketitle

\section{Introduction}\label{sec:intro}

Big-data analytics is an ubiquitous practice with a noticeable impact on our lives. Our digital interactions produce massive amounts of data that are analyzed in order to discover unknown patterns or correlations, which help us draw safer conclusions or make informed decisions. At the core of this lies Machine Learning (ML) for devising complex data models and predictive algorithms that provide hidden insights or automated actions, while optimizing certain objectives.  Example applications that successfully employ ML are market forecast, service personalization, speech/face recognition, autonomous driving, health diagnostics and security analytics.

Of course, data analysis is only as good as the analyzed data, but this goes beyond the need to properly inspect, cleanse or transform high-fidelity data prior to its modeling: In most learning domains, analyzing ``big data'' is of twofold semantics: \emph{volume} and \emph{variety}.

First, the larger the dataset available to an ML algorithm, the better its learning accuracy, as irregularities due to outliers fade away  faster. Indeed, scalability to large dataset sizes is very important, especially so in unsupervised learning, where model inference uses unlabelled observations (evading points of saturation, encountered in supervised learning, where new training sets improve accuracy only marginally). Also, the more varied the collected data, the more elaborate its analysis, as degradation due to noise reduces and domain coverage increases. Indeed, for a given learning objective, say classification or anomaly detection, combining more datasets of similar type but different origin enables discovery of more complex, interesting, hidden structures and of richer association rules (correlation or causality) among  attributes.
So, ML models improve their predictive power when they are built over multiple datasets owned and contributed by different entities, in what is termed \emph{collaborative learning}---and widely considered as the golden standard~\cite{BerkeleyAIView}.

\smallskip \noindent\textbf{Privacy-preserving hierarchical clustering.}  Several learning tasks of interest, across a variety of application domains, such as healthcare or security analytics, demand deriving accurate ML models over highly sensitive data---e.g., personal, proprietary, customer, or other types of data that induce liability risks. By default, since collaborative learning inherently implies some form of data sharing, entities in possession of such confidential datasets are left with no other option than simply running their own local models, severely impacting the efficacy of the learning task at hand. Thus, 
\emph{privacy risks are the main impediment to collaboratively learning} richer models over large volumes of varied, individually contributed, data.

The security and ML community  has embraced the concept of \emph{Privacy-preserving Collaborative Learning} (PCL), the premise being that effective analytics over sensitive data is feasible by building global models in ways that protect privacy. \textcolor{black}{This is closely related to (privacy-preserving) ML-as-a-Service~\cite{ppmlaas,CryptoNets,crip,crop} that utilizes cloud providers for ML tasks, without parties revealing their sensitive raw data (e.g., using encrypted or sanitized data.}
Existing work on PCL {mostly focuses on supervised rather than unsupervised learning tasks} (with a few exceptions such as $k$-means clustering).  As unsupervised learning is a prevalent paradigm, the design of ML protocols that promote collaboration and privacy is vital.

In this paper, we study the problem of \emph{privacy-preserving hierarchical clustering}. \textcolor{black}{This unsupervised learning method  groups data points into similarity clusters, using some well-defined distance metric. The ``hierarchic'' part is because  each data point starts  as a separate ``singleton'' cluster and clusters are iteratively merged building increasingly larger clusters. This process forms a natural hierarchy of clusters that is part of the output, showing how the final clustering was produced.} We present scalable cryptographic protocols that allow two parties to privately learn a model for the joint clusters of their combined datasets. Importantly, we propose a formal security definition for this task in the MPC framework  and prove our protocols satisfy it. In contrast, prior works for privacy-preserving hierarchical clustering have proposed crypto-assisted protocols but without offering rigorous security definitions or analysis (e.g.,~\cite{pphc3,PHC14,pphc5}; see detailed discussion in Section~\ref{sec:related}).

\smallskip\noindent\textbf{Motivating applications.} Hierarchical clustering is a class of unsupervised learning methods that build a \emph{hierarchy of clusters} over an input dataset, typically in bottom-up fashion.  Clusters are initialized to each contain a single input point and are iteratively merged in pairs, according to a \emph{linkage} metric that measures clusters' closeness based on their contained points. Here, unlike other clustering methods ($k$-means or spectral clustering), different distance metrics can define cluster linkage (e.g., nearest neighbor and diameter for \emph{single} and \emph{complete} linkage, respectively) and flexible conditions on these metrics can determine when merging ends. The final output is a \emph{dendrogram} with all formed clusters and their merging history. This richer clustering type is widely used in practice, often in areas where the need for scalable PCL solutions is profound.

In healthcare, for instance, hierarchical clustering allows researchers, clinicians and policy makers to process medical data and discover useful correlations to improve health practices---e.g., discover similar genes types~\cite{Eisen98}, patient profiles most in need of targeted intervention~\cite{Weir00,Newcomer11} or changes in healthcare costs for specific treatments~\cite{Liao16}. To be of any predictive value, such  data contains sensitive information (e.g., patient records, gene information, or PII) that must be protected, also due to legislations such as HIPPA in US or GDPR in~EU.
Also, in security analytics, hierarchical clustering allows enterprise security personnel to process log data on network/users activity to discover suspicious or malicious events---e.g., detect botnets~\cite{BotMiner}, malicious traffic~\cite{ExecScent}, compromised accounts~\cite{Cao14}, or malware~\cite{Bayer09}. Again, such data contains sensitive information (e.g., employee/customer data, enterprise security posture, defense practices, etc.) that must be protected, also due to industrial regulations or for reduced liability. As such, without privacy provisions for joint cluster analysis, entities are restricted to learn only local clusters, thus confined in accuracy and effectiveness.  E.g., a clinical-trial analysis over patients of one hospital may introduce bias on geographic population, or network inspection of one enterprise may miss crucial insight from attacks against others.

In contrast, our treatment of clustering as a PCL instance is a solid step towards richer classification. Our protocols for private hierarchical clustering incentivize entities to contribute their private datasets for joint cluster analysis over larger and more varied data collections, to get in return more refined results. For instance, hospitals can jointly cluster medical data extracted from their combined patient records, to provide better treatment, and enterprises can jointly cluster threat indicators collected from their combined SIEM tools, to present timely and stronger defenses against attacks.\footnote{In line with current trends toward collaborative learning in healthcare/security analytics; e.g., AI-based clinical-trial predictions~\cite{AITrials}, threat-intelligence sharing~\cite{ThreatIntelligence,CTA,OTX,FBTX}.}  At all times, data owners protect the confidentiality of their private data and remain compliant with current regulations.

\smallskip \noindent\textbf{Challenges and insights.} A first challenge we faced is how to  rigorously specify the secure functionality that such protocols must achieve. A secure protocol guarantees that no party learns anything about the input of the other party, \emph{except what can be inferred after parties learn the output}. But since the output dendrogram  of hierarchical clustering already includes the (now partitioned) input, this problem cannot directly benefit from MPC. This issue is partially the reason why previous approaches for hierarchical clustering (see discussion in Section~\ref{sec:related} and an excellent survey of related work by Hegde et al.~\cite{hegde}) lack formal security analysis or have significant information leakage. To overcome this, our approach is 
 to modify and refine what private hierarchical clustering should produce, redacting the joint output---sufficiently enough, to allow the needed input privacy protection, but minimally so, to preserve the learning utility. We introduce a security notion that is based on \emph{point-agnostic dendrograms}, which explicitly capture only the merging history of formed joint clusters and useful statistics thereof, to  balance the intended accuracy against the achieved privacy. To the best of our knowledge, our formal security definition (Section~\ref{sec:def}) is the first such attempt for the case of hierarchical clustering.
 
The next challenge is to securely realize this functionality efficiently. Standard tools for secure two-party computation, e.g., garbled circuits~\cite{Yao1,Yao2}, result in large communication, while fully homomorphic encryption~\cite{FHE} is still rather impractical, so designing scalable hierarchical clustering  PCL protocols is challenging. Moreover, hierarchical clustering of $n$ points is already computation-heavy---of $\mathcal{O}(n^3)$ cost. As such, approximation algorithms, e.g., CURE~\cite{cure}, are the \emph{de facto}  means to scale to massive datasets, but incorporating approximation to private computation is not trivial---as complications often arise in defining security~\cite{approx1}. 

In Section~\ref{sec:main}, we follow a modular design approach and use cryptography judiciously by devising our main construction as a \emph{mixed protocol} (e.g.,~\cite{Tasty,ABY,mixed3}). We decompose our refined hierarchical clustering into building blocks and then we select a combination of tools that achieves fast computation and low bandwidth usage. In particular, we conveniently use garbled circuits for cluster merging, but additive homomorphic encryption~\cite{Paillier99} for cluster encoding, while securely ``connecting'' the two steps' outputs.

In Section~\ref{sec:analysis}, we evaluate the performance and security of our main protocol and present an optimized variant of $O(n^2)$ cost for single linkage. In Section~\ref{sec:approx}, we  integrate the CURE method~\cite{cure} for \emph{approximate clustering} into our design, to get the best-of-two-worlds quality of high scalability and privacy. We study different secure approximate variants that exhibit trade-offs between efficiency and accuracy without extra leakage due to approximation. 
In Section~\ref{sec:eval}, we report results from the experimental evaluation of our protocols on synthetic and real data that confirm their practicality. For example, end-to-end execution of our private approximate single-linkage protocol for $1$M $10-d$ records, achieves $97.09\%$ accuracy at very with only $35$sec of computation time.
%
%

   \vspace{.1cm}
   \noindent\textbf{Summary of contributions.}  Overall, in this work  our results can be summarized as follows: 
   \begin{itemize}[leftmargin=*]

  \item We provide a formal definition and secure two-party protocols for private hierarchical clustering for single or complete linkage.

  \item We present an optimized protocol for single linkage that significantly improves the computational and communication costs.

  \item We combine approximate clustering methods with our protocols to get variants that achieve both scalability and strong privacy.

  \item We experimentally evaluate the performance of our protocols via a prototype implementation over synthetic and real datasets.

 \end{itemize}


\section{Preliminaries}
\label{sec:prelims}

\noindent {\bf Hierarchical clustering (HC).} 
%
\textcolor{black}{
For fixed positive integers $d,l$, let $\dataset = \{v_i|v_i \in \mathbb{Z}_{2^l}^d\}_{i=1}^n$ be an \emph{unlabeled} indexed dataset of $n$ $d$-dimensional points, where w.l.o.g, we set the domain to  $\{0,\ldots,2^l -1\}$.
}
Over pairs $x,y\in\dataset$ of points, \emph{point distance} is measured using the standard \emph{square Euclidean distance} metric $\dist(x,y) = \sum_{j=1}^d (x_j -  y_j)^2$. Over pairs $X, Y\subseteq \dataset$ of sets of points, \emph{set closeness} is measured using a \emph{linkage distance} metric $\delta(X,Y)$, as a function of the cross-set distances of points contained in $X$, $Y$. The most commonly used linkage distances are the \emph{single linkage} (or nearest neighbor) defined as $\delta(X,Y) = \min_{x \in X, y \in Y} \dist(x,y)$, and the \emph{complete linkage} (or diameter) defined as $\delta(X,Y) = \max_{x \in X, y \in Y} \dist(x,y)$.

\setlength{\floatsep}{3pt}
\setlength{\textfloatsep}{3pt}

Standard agglomerative HC methods use set closeness to form clusters in a bottom-up fashion, as described in algorithm $\mathsf{HCAlg}$ (Figure~\ref{f:hc}). It receives an $n$-point dataset $\dataset$ and groups its points into a total of $\ell_t\leq n$ \emph{target} clusters, by iteratively merging pairs of closest clusters into their union. The merging history is stored (redundantly) in a \emph{dendrogram} $T$, that is, a forest of clusters of $n-\ell_t+1$ levels, where siblings correspond to merged clusters and levels to dataset partitions, build level-by-level as follows:

\begin{itemize}[leftmargin=10pt]

\item Initially, each input point $v_i \in \dataset$ forms a singleton cluster $\{ v_i \}$ as a leaf in~$T$ (at its lowest level~$n$).

\item Iteratively, in $n-\ell_t$ \emph{clustering rounds}, the $i$ root clusters (at top level $i$) form $i-1$ new root clusters in $T$ (at higher level $i-1$), with the closest two merged into a union cluster as their parent, and each other cluster copied to level $i-1$ as its parent.

\end {itemize}

\noindent When a new level of $\ell_t$ target clusters is reached, $\mathsf{HCAlg}$ halts and outputs $T$. The exact value of $\ell_t\in [1:n]$ is determined during execution via a predefined condition $\mathsf{End}$ checked over the current state $T$ and a termination parameter $t$ provided as additional input. This allows for flexible termination conditions---e.g., stopping when inter-cluster distance drops below an  threshold specified by $t$, or simply when exactly $\ell_t=t$ target clusters are formed.

Typically, the dendrogram $T$ is augmented to store some associated cluster metadata, by keeping, after any union/copy cluster is formed, some useful statistics over its contained points. Common such statistics for cluster $C$ is its \emph{size} $size(C) = |C|$ and \emph{representative} value $rep(C)$, usually defined as its \emph{centroid} (i.e., a certain type of average) point.
Overall, for a set $M$ of cluster statistics of interest and specified linkage distance and termination condition, $\mathsf{HCAlg}$ is viewed to operate on indexed dataset $\dataset = \{ v_i\}_{i=1}^n$ and return an \emph{$M$-augmented dendrogram $T$}, comprised of: (1) the \emph{forest structure of dendrogram} $T$, specifying the full merging history of input points into formed clusters (from $n$ singletons to $\ell_t$ target ones); (2) the \emph{cluster set} $C(T)$; and (3) the \emph{metadata set} $M(T)$ associated with (clusters in)~$T$. Assuming that $\mathsf{HCAlg}$ employs a fixed tie-breaking method in merging clusters, its execution is \emph{deterministic}.

\begin{figure}[t!]
\begin{minipage}[b]{6.5in}
	\fbox{
           \small
		\begin{minipage}{.5\textwidth}
                  \noindent\textbf{Hierarchical Clustering Algorithm $\mathsf{HCAlg}$}

                  \smallskip\textbf{Input:} Indexed set $\dataset = \{{\bf v}_i\}_{i=1}^n$, termination parameter~$t$

                  \smallskip\textbf{Output:} Dendrogram $T$, clusters $C(T)$, metadata $M(T)$
                  
                  \smallskip\textbf{Parameters:} Linkage distance $\delta(\cdot,\cdot)$, termination condition $\mathsf{End}(\cdot,\cdot)$, cluster statistics set $M\supseteq \{rep(\cdot),size(\cdot)\}$
                  
                \smallskip\textbf{[Initially, at level $n$]}



               \smallskip~~1. {\bf Initialize dendrogram $T$:} For each $i = 1, \ldots, n$:

                 ~~~~~-- Create node $u_i$ as the $i$th left-most leaf in $T$.

                 ~~~~~-- Set $C(u_i) = \{{\bf v}_i\}$ as the singleton cluster of~$u_i$.

                  ~~~~~-- Compute $M(u_i) = \{m({\bf v}_i)|m\in M\} $ as statistics of~$u_i$.
                  
                  \smallskip~~2. {\bf Set up linkages:} Compute linkages of all pairs of 
                  
                  ~singleton clusters as a dictionary $D$, where $\{C(u_i),C(u_j)\}$ 
                  
                  ~is keyed under $\delta(C(u_i),C(u_j))$, $1\leq i < j \leq n$.

                    \smallskip\textbf{[Iteratively, at level $i = n, \ldots, \ell_t + 1$]}


                   \smallskip~~1. {\bf Update $T$:} If $N_i$ is the set of nodes in $T$ at level $i$:

                   ~~~~~-- Find in $D$ the min-linkage pair $(u,u')$ of nodes in~$N_i$,

                   ~~~~~breaking ties using a fixed rule over leaf-node indices.

                   ~~~~~-- Create node $w\in N_{i-1}$ as parent of $u$ and $u'$; set 
                   
                   ~~~~~$C(w) = C(u)\cup C(u')$; for each node $\bar{u} \in N_i-\{u,u'\}$,

                   ~~~~~create node $\bar{w}\in N_{i-1}$ as parent of $\bar{u}$; set $C(\bar{w}) = C(\bar{u})$.

                      ~~~~~-- For each node $\hat{w}\in N_{i-1}$, compute $M(\hat{w})$.

                      \smallskip~~2. {\bf Check termination:} If $\mathsf{End}(T,t)==1$, terminate.

                      \smallskip~~3. {\bf Update linkages:} Compute linkage $\delta(C(w),C(\bar{w}))$, for

                      ~~~all $\bar{w}\in N_{i-1}-\{w\}$, and consistently update dictionary~$D$.
	
		\end{minipage}
	}
      \end{minipage}
\vspace{-0.4cm}
\caption{Agglomerative hierarchical clustering.}
\label{f:hc}
\end{figure}

\smallskip\noindent {\bf Secure computation and threat model.} We consider the standard setting for private two-party computation, where two parties wishing to evaluate function $f(\cdot,\cdot)$ on their individual, private inputs $x_1$, $x_2$, engage in an interactive cryptographic protocol that upon termination returns to them the common output $y = f(x_1,x_2)$. Protocol security has this semantics: Subject to certain computational assumptions and misbehavior types during protocol execution, no party learns anything about the input of the other party, other than what can be inferred by its own input $x_i$ and the learned result $y$. In this context, we study privacy-preserving hierarchical clustering in the \emph{semi-honest} adversarial model which assumes that parties are honest, but curious: They will follow the prescribed protocol but also seek to infer information about the input of the other party, by examining the transcript of exchanged messages---the latter, assumed to be transferred  over a reliable channel.

\textcolor{black}{Although, in practice, parties may choose to be malicious, deviating from the prescribed protocol if they can benefit from this and can avoid detection, the semi-honest adversarial model still has its merits, especially in the studied PCL setting. 
Namely, it provides essential privacy protection for any privacy-aware party to enter the joint computation to benefit from collaborative learning. We note that, by trading off efficiency, security can be  hardened via known generic techniques for compiling protocols secure in this model into counterparts secure against malicious parties.}

\smallskip\noindent {\bf Garbled circuits}. \iffull  One of the most widely used tools for two-party secure computation, \fi \emph{Garbled Circuits} (GC)~\cite{Yao1,Yao2} allow two parties to evaluate a boolean circuit on their joint data without revealing their respective inputs. This is done by generating an encrypted truth table for each gate while evaluating the circuit by decrypting these tables in a way that preserves input privacy. \iffull In Appendix~\ref{app:gc}, we provide more details about the GC framework.\fi

\smallskip\noindent {\bf Homomorphic encryption.}
This technique allows carrying out operations over  encrypted data. Fully Homomorphic Encryption (FHE)~\cite{FHE} can evaluate arbitrary functions over ciphertexts, but remains rather impractical. Partially homomorphic encryption supports only specific arithmetic operations over ciphertexts, but allows for very efficient implementations~\cite{rsa,Paillier99}. We use Paillier’s scheme for Additively Homomorphic Encryption (AHE)~\cite{Paillier99}, summarized as follows. For security parameter $\lambda$, keys generated by running $(\pk, \sk) \larr \gen(1^\lambda)$ and a public RSA modulus $N$, the scheme encrypts (with public key $\pk$) any message $m$ in the plaintext space $\Z_N$ into a  ciphertext $\en{m}$, ensuring that decryption (with secret key $sk$) of any ciphertext product $\en{m}\cdot\en{m'}\mod N^2$ (computable without $\sk$) results in the plaintext sum $m+m' \mod N$. Thus, decrypting $\en{m}^k \mod N^2$ results in $km \mod N$, and the ciphertext product $\en{m}\cdot\en{0}$ results in a fresh encryption of $m$. 


\section{Formal Problem Specification}
\label{sec:def}

We introduce a  model for studying private hierarchical clustering, the first to provide formal specifications for secure two-party protocols for this central PCL problem. Importantly, we define security for a refined learning task that achieves a meaningful balance between the intended accuracy and privacy---a necessary compromise for the problem at hand to even be defined as a PCL instance!

We first formulate two-party privacy-preserving hierarchical clustering as a secure computation. Parties $\Pone$, $\Ptwo$ hold independently owned datasets $P$, $Q$ of points in $\mathbb{Z}_{2^l}^d$, and wish to perform a collaborative hierarchical clustering over the combined set $\dataset = P \cup Q$. They agree on the \emph{exact specification} $f_{HC}$ of this learning task, as a function of their individually contributed datasets that encompasses all other parameters (e.g., for termination).

Let $\Pi$ be a two-party protocol that \emph{correctly} realizes $f_{HC}(\cdot,\cdot)$: Run jointly on inputs $x_1$, $x_2$, $\Pi$ returns the common output~$f_{HC}(x_1,x_2)$.  Thus, parties $\Pone$, $\Ptwo $ can learn cluster model $f_{HC}(P,Q)$ by running protocol $\Pi$ on their inputs $P$, $Q$. As discussed, $\Pi$ is considered to be secure if its execution prevents an honest-but-curious party from learning anything about the other party's input that is not implied by the learned output. We formalize this intuitive privacy requirement via the standard two-party \emph{ideal/real world} paradigm~\cite{GMW87}.

\smallskip\noindent {\bf Ideal functionality.} First, we define what one can best hope for.
Cluster analysis with \emph{perfect} privacy is trivial in an \emph{ideal} world, where $\Pone$, $\Ptwo $ instantly hand-in their inputs $x_1$, $x_2$ to a \emph{trusted} third party, called the \emph{ideal functionality $f_{HC}$}, that computes and announces $f_{HC}(x_1,x_2)$ (and explodes). Here, the use of terms ``perfect'' and ``ideal'' is fully justified for no information about any private input is leaked \emph{during} the computation. Some information about $x_1$ or $x_2$ may be inferred \emph{after} the output is announced, by combining the known $x_2$ or $x_1$ with the learned $f_{HC}(x_1,x_2)$: It is the \emph{inherent price for collaboratively learning a non-trivial function}.

In the \emph{real} world, $\Pone$, $\Ptwo $ learn $f_{HC}(x_1,x_2)$ by interacting in the joint execution of a protocol $\Pi$. We measure the privacy quality of $\Pi$ against the ideal-world perfect privacy, dictating that running $\Pi$ is effectively \emph{equivalent} to calling the ideal functionality $f_{HC}$. Informally, $\Pi$ \emph{securely realizes} $f_{HC}$, if anything computable by an efficient semi-honest\ party $\Pany$ in the real world, can be simulated by an efficient algorithm (called the simulator $\simu$), acting as $\Pany$ in the ideal world; i.e., $\Pi$ leaks no information about a private input during execution, subject to the price for learning~$f_{HC}(x_1,x_2)$.

Next comes the question of which ideal functionality $f_{HC}$ should $\Pi$ securely realize for private joint hierarchical clustering? Though tempting, equating $f_{HC}$ with the legacy algorithm $\mathsf{HCAlg}$ (Figure~\ref{f:hc}), thus learning a full-form augmented dendrogram, slides us into a degeneracy. Assume  $f_{HC}$ merely runs $\mathsf{HCAlg}$ on the combined indexed set $\dataset = P \cup Q=\{d_k\}_{k=1}^{n}$, $n=|P|+|Q|$.\footnote{If $P$, $Q$ are indexed, then $\dataset =Q\|P$, or else a \emph{fixed} ordering is used.} The learned model is the dendrogram $T$ along with its associated clusters $C(T)$ and metadata $M(T)$. But set $C(T)$ itself reveals the input $\dataset$; in this case, the price for collaborative learning is full disclosure of sensitive data and nothing is to be protected! \textcolor{black}{This raises the question of limiting exactly what information about $P,Q$ should be revealed by $f_{HC}$ which is the focus of the remainder of this section.}

\begin{figure}[t!]
\begin{minipage}[b]{6.5in}
	\fbox{
		\small
          \begin{minipage}{.5\textwidth}
                  \noindent\textbf{Ideal Functionality $f_{HC}^*(\cdot, \cdot)$}

                  \smallskip\textbf{Input:} Sets $P=\{p_i\}_{1}^{n_1}$, $Q=\{q_j\}_{1}^{n_2}$

                  \smallskip\textbf{Output:} Dendrogram $T^*$, metadata $M^*\supseteq \{rep(\cdot),size(\cdot)\}$
                  
                  \smallskip\textbf{Parameters:} Linkage distance $\delta(\cdot,\cdot)$, termination condition $\mathsf{End}(\cdot,t)$, cluster statistics set $M$, selection function $\mathsf{S(\cdot)}$
                  
                  \smallskip\textbf{[Pre-process]} Form input of size $n=n_1+n_2$ for  $\mathsf{HCAlg}$:


                   \smallskip~~1. Set $\dataset=\{d_k\}_1^n$ s.t. $d_k=p_k$, if $k\leq n_1$, or else $d_k=q_{k-n_1}$.
                   
                  \smallskip~~2. Pick random permutation $\pi:[n]\rightarrow [n]$; set $\dataset^*=\pi(\dataset)$.


                    \smallskip\textbf{[HC-process]} Run $\mathsf{HCAlg}(\dataset^*,t)$ w/ parameters $\delta$, $M$, $\mathsf{End}$.

                \smallskip\textbf{[Post-process]} Redact output $T^*$, $C(T^*)$, $M(T^*)$ of  $\mathsf{HCAlg}$:
             

                \smallskip~~1. Set $M^*=\emptyset$; $\forall v\in T^*$: if $\mathsf{S}(v)==1$, $M^*\leftarrow M^*\cup \{M(v)\}$.
                
                \smallskip~~2. Return $T^*$,  $M^*$.

		\end{minipage}
	}
      \end{minipage}
\vspace{-0.3cm}
\caption{Ideal functionality $f_{HC}^*$ for hierarchical clustering.}
\label{fig:idealhc}
\end{figure}

\smallskip\noindent {\bf Refined cluster analysis.}  In the PCL setting, we need a new definition of hierarchical clustering that distills the full augmented dendrogram $\{T, C(T),M(T)\}$ into a redacted, but still useful, learned model, balancing between accuracy (to benefit from clustering) and privacy (to allow collaboration). If allowing the ideal functionality $f_{HC}$ to return  $C(T)$ is one extreme that diminishes privacy, removing the dendrogram $T$ from the output---to learn only about its associated information $C(T)$, $M(T)$---is another that diminishes accuracy. Indeed, if $T$, which captures the full merging history in its structure, is excluded from the output of $f_{HC}$, a core feature in HC is lost: the ability to gain insights on \emph{how target clusters were formed}, under what hierarchies and in which order. This renders the HC analysis only as good as much simpler techniques (e.g., $k$-means) that merely discover pure similarity statistics of target clusters. As the motivation for studying collaborative HC as a prominent and widely used unsupervised learning task, in the first place, lies exactly on its ability to discover such rich inter-cluster relations, we must keep the forest structure of $T$ in $f_{HC}$'s output.\footnote{Cluster hierarchy is vital in HC learning, e.g., in healthcare, revealing useful causal factors that contribute to prevalence of diseases~\cite{Eisen98} and in biology, revealing useful relationships among plants, animals and their habitat ecological subsystems~\cite{Girvan2002Community}.} 

Avoiding the above two degenerate extremes suggests that the learned model $f_{HC}(P,Q)$ should necessarily include the cluster hierarchy $T$ but not the clusters $C(T)$ themselves. Yet, the obvious \emph{middle-point} approach of learning model $f^m_{HC}(P,Q) =\{T, M(T)\}$ remains suboptimal in terms of privacy protections, as the learned output can still be strongly correlated to exact input points. Indeed, given $T$ and a party's own input, inferring points of the other party's input simply amounts to identifying singleton clusters, which is generally possible by inspecting and correlating the (hard-coded in $\mathsf{HCAlg}$) indices in~$\dataset$ with the metadata associated to singletons (or their close neighbors). For instance, if $w$ is the parent of singleton $u$ and cluster $u'$ in $T$, then $\Pone$ can infer input point $C(u)$ of $\Ptwo$, either directly from output $M(u)$, if $u$ is known to store none of its input points, or indirectly from $M(u')$, $M(w)$, if these output values imply a value of $M(u)$ that is consistent with none of its own inputs. 

Also, even without singleton clusters in the output, there is still leakage from the positioning of the points at the leaf level of $T$. E.g., assuming $P,Q$ are ordered from left to right, a merging of two points at the right half of the tree  during the first merge reveals to  $\Pone$ that  $\Ptwo$ has a pair of points with smaller distance than the minimum distance observed among points in $P$. Hence, it is  crucial to eliminate  information about the  positioning of  clusters in $T$.

\smallskip\noindent {\bf Point-agnostic dendrogram.} Such considerations naturally lead to a new goal: We seek to refine further, but minimally so, the middle-point model $f^m_{HC}(P,Q) =\{T, M(T)\}$ into an optimized model $f^*_{HC}(P,Q)=\{T^*, M^*(T)\}$, whereby \emph{no private input points directly leak to any of the parties}, after the output is announced. This quality is well-defined, intuitive and useful: Unless the intended joint hierarchical clustering explicitly copies some of input points to the output, the learned model $f^*_{HC}(P,Q)$ should allow no party to explicitly learn, that is, to deterministically deduce with certainty, any of the unknown input points of the other party.

We accordingly set our ideal functionality $f^*_{HC}$ for hierarchical clustering to outputs a \emph{point-agnostic augmented dendrogram}, defined by merely running algorithm $\mathsf{HCAlg}$, subject to a twofold \emph{correction} of its input $P$, $Q$ and returned dendrogram (Figure~\ref{fig:idealhc}):

\begin{itemize}[leftmargin=10pt]
\item {\bf Pre-process input:} Run $\mathsf{HCAlg}$ on indexed set $\dataset^*$ that is a \emph{random permutation} over the combined set  $\dataset = P \cup Q = \{d_k\}_{k=1}^{n}$.

\item {\bf Post-process output:} Return the output $T^*$, $C(T^*)$, $M(T^*)$ of $\mathsf{HCAlg}$ \emph{redacted} as $T^*,M^*(T^*)\subset M(T^*)$, including  metadata of only a few \emph{safe} clusters in~$T^*$.

\end {itemize}

\noindent Our ideal functionality $f^*_{HC}$ refines  the ordinary dendrogram $T$, $C(T)$, $M(T)$: Running $\mathsf{HCAlg}$ on the randomly permuted input $\dataset^*$ (instead of $\dataset$) results in a new \emph{randomized} forest structure $T^*$ (instead of $T$) and, although its associated sets of formed clusters $C(T^*)$ and metadata $M(T^*)$ remain the same, the learned model includes no elements from $C(T^*)$, but only specific elements from $M(T^*)$, determined by a \emph{selection function} $\mathsf{S}(\cdot)$ (as a parameter agreed upon among the parties and hard-coded in $f^*_{HC}$).  Such metadata is safe to learn, in the sense that it does not directly leak any  input points.

We propose the following two orthogonal strategies for safe metadata selection for point-agnostic dendrograms:
\begin{itemize}[leftmargin=10pt]

\item {\bf $s$-Merging selection:}  $M(w) \in M(T^*)$ if $w$ is the parent of $u$, $u'$ in $T^*$ and $|C(u)|, |C(u')|> s$: any non-singleton cluster formed by merging two clusters of size above  threshold~$s>0$, is safe;

\item {\bf Target selection:}  $M(w) \in M(T^*)$ if $w$ is root in $T^*$: any target cluster at level $\ell_t$ in $T^*$ is safe.

\end {itemize}

\noindent Above, the first strategy ensures that no direct leakage of private input points occurs by correlating statistics of thin neighboring clusters; in particular, no cluster statistics are learned for singletons or their parents ($s=1$), thus eliminating the type of leakage allowed by model $f^m_{HC}(P,Q)$. The second strategy ensures that only statistics of target clusters are learned, that is, input points may be directly learned only explicitly as part of the intended cluster analysis.

Overall, the resulting dendrogram is point-agnostic in the sense that neither the forest structure of $T^*$ nor the  metadata $M(T^*)$   reveal which singletons a party's points are mapped to. As points are {randomly} mapped to singletons, ties in cluster merging are {randomly} broken, and no statistics are learned for singleton (or thin) clusters, no party can deduce with certainty any of the other party's input points. \textcolor{black}{For instance, the applied permutation eliminates leakage from the positioning of the singleton cluster at the leaves that, in our previous example, allowed one to infer whether the other party owned points with smaller distance than its own pairs, from the first-round clustering result. More generally, anything inferable about a party's private input relates to a {meta-analysis} that must necessarily encompass the (unknown) input distribution and the  random permutation used by~$f^*_{HC}$.} This can be viewed as an \emph{inherent price of collaborative hierarchical clustering}.
\Comment{
The resulted dendrogram is point-agnostic, at least, in the sense that neither the forest structure of $T^*$ nor any learned metadata can, by itself or in combination, reveal which singletons in $T^*$ the input points of a party map to. Indeed, since input points are \emph{randomly} mapped to singleton nodes, ties in cluster merging are also \emph{randomly} broken, and since no cluster statistics are learned for singletons, \emph{no party can deterministically deduce with certainty} any of the input points of the other party. As a worst case example, learning the average value of a cluster of size 2 cannot directly leak an input point of the other party, as this cluster can be possibly formed by points of either party or of both parties.

Anything that can be possibly inferred about the other party's input corresponds to a \emph{probabilistic meta-analysis} that must necessarily encompass an assumed input distribution of the other party's input and the underlying unknown random permutation used by~$f^*_{HC}$. Such possible inference is viewed as the price of collaborative HC-based cluster analysis. Back to the example of learning the average value of a cluster of size 2, a party may guess with higher probability that this cluster is formed by its own input points, which in turn can slightly revise the probability distribution that is assumed for the input points of the other party, but such indirect inference is compensated by the assumption that there some value in analyzing this exact cluster, through the learned average value. We note, however, that this inherent price of collaborative cluster analysis can be traded with some loss in the intended accuracy, by further redacting from the returned set $M(T^*)$ the statistics of certain type of nodes in~$T^*$. We provide two such additional refinements: 

\begin{itemize}

\item {\bf Thin-cluster pruning:}  $M(T^*)$ excludes the statistics of clusters of size at most a given threshold value~$s$.

\item {\bf Non-target cluster pruning:}  $M(T^*)$ excludes the statistics of all non-target clusters.

\end {itemize}

Our ideal functionality $f^*_{HC}$ is detailed in~Figure~\ref{fig:idealhc}, encompassing any of the three discussed pruning methods (i.e., excluding statistics of singleton, thin or non-target clusters) via an extra parameter, a function that selects any cluster $C$ whose metadata $M(C)$ is included to the final output, where $M(C)$ includes its representative $rep(C)$ and size $size(C)$.
}The following  defines the security of  privacy-preserving hierarchical clustering.

\vspace{-.2cm}
\begin{definition}\label{def:sec} A two-party protocol $\Pi$, jointly run by $\Pone$, $\Ptwo$ on respective inputs $x_1$, $x_2$ using individual random tapes $r_1$, $r_2$ that result in incoming-message transcripts $t_1$, $t_2$, is said to be \emph{secure} for \emph{collaborative privacy-preserving hierarchical clustering} in the presence of {\em static, semi-honest adversaries}, if it securely realizes the ideal functionality $f^*_{HC}$ defined in Figure~\ref{fig:idealhc}, by satisfying the following: For $i=1,2$ and for any security parameter $\secp$, there exists a non-uniform probabilistic polynomial-time simulator $\simu_{\Pany}$ so that $\simu_{\Pany}(1^{\secp}, x_i, f^*_{HC}(x_1,x_2)) \cong \view_{\adv^{\Pi}_{\Pany}}\triangleq\{r_i,t_i\}$.
\end{definition}

\Comment{
\smallskip\noindent\textbf{Security definition.} We are ready to provide the definition for privacy-preserving hierarchical clustering.

\begin{definition}\label{def:func} A protocol $\pi=\langle \Pone, \Ptwo\rangle$ is a \emph{secure privacy-preserving hierarchical clustering} in the presence of {\em static, semi-honest adversaries}  if it realizes the ideal functionality $f^*_{HC}$ (defined in Figure~\ref{fig:idealhc}). Specifically, for $i=1,2$ and for all $\secp$, there exists non-uniform probabilistic polynomial-time $\simu_{\Pany}$ such that $\simu_{\Pany}(1^{\secp}, x_i, f^*_{HC}(x_1,x_2)) \cong \view_{\adv^{\pi}_{\Pany}}$, where $x_1,x_2$ are the respective inputs of $\Pone,\Ptwo$, and $\view_{\adv^{\pi}_{\Pany}}$ consists of the randomness tape of $\Pany$ and all incoming messages received while running protocol $\pi$.
\end{definition}
}
\Comment{
\begin{figure}[tbp]
	\centering
	\includegraphics[scale=0.2]{dendrogram.pdf}
	\caption{(Left) Dendrogram produced by hierarchical clustering over lists $P=(p_1,p_2,p_3)$ and $Q=(q_1,q_2,q_3)$ for $3$ target clusters. (Right) The dendrogram after permuting the concatenated lists under permutation $\pi$.}\label{f:dendrogram}
		\vspace{-0.3cm}
\end{figure}
}

\ignore{
\alina{Move this paragraph to Key Insights}
Similar to a line of recent work in secure computation~\cite{ABY,popa,secureML,opt3,opt7}, we leverage \emph{mixed protocols}, a design paradigm that combines AHE with garbled circuits. We decompose the computation into individual building blocks (e.g., comparison, minimum distance computation, minimum cluster linkage), design secure protocols for each of these with the most efficient secure computation method, and finally combine the components into a secure end-to-end protocol for two-party secure hierarchical clustering.}

\Comment{

First, during the dendrogram initialization, the left-to-right ordering of singleton clusters at level $n$ comprise a \emph{random permutation} of the input 
set $\dataset = P \cup Q$ of the HC algorithm. Second, metadata of singleton nodes is removed from $T'$

\ignore{
\noindent\textbf{\underline{Attempt 2: Removing clusters from lower layers.}} A  different approach would be to remove the cluster $\C(v)$ from each node $v$ and only output the dendrogram $T$'s structure and the representative $rep(v_1).\dots,rep(v_{\ell_t})$ and sizes $|\C(v_1)|,\dots, |\C(v_{\ell_t})|$ for each of the final ${\ell_t}$ clusters. At first sight, it seems that this approach is sufficient. It does reveal more than $k$-means due to the dendrogram structure, but it does not directly reveal the other party's inputs. However, we believe that the information leaked is still too much, according to the following example. Let $u,v$ be the two nodes that correspond to the merged cluster in the first iteration. If both $u,v$ correspond to points from $Q$ (which can be easily deduced by checking the indexes of $u,v$ in level $n$), then the first party learns that the other party holds two points that have smaller distance than any of his points. This example trivially generalizes to higher layers in the dendrogram, therefore we are motivated to reduce this leakage.}


\noindent\textbf{\underline{Attempt 2: Keep only the dendrogram structure.}}  A second approach is to keep only the dendrogram structure without the intermediate clusters $\C(v)$ at layers $\ell_t+1,\dots,n$. This has the advantage of protecting the inputs at the last layer. However, it will reveal the exact mapping between input points and the final clusters, which results in considerable leakage. For instance, a party can infer upper bounds on distances between its points and points owned by the other party, or distances between the other party's points since it knows the exact order in which clusters were merged.

Finally, we propose to keep the dendrogram structure, but  permute nodes only once at the last layer. We believe that this represents the right balance between preserving the utility of the dendrogram, and reducing the amount of  revealed information. The parties can still infer in which order the clusters where merged  and the internal topology of the clusters, which has applications in medicine and biology~\cite{Eisen98,Girvan2002Community}. At the same time, this approach hides the mapping of points to clusters and the exact ordering of cluster merging, thus preventing simple attacks based on inferring distances between points.

\Comment{

To overcome this obstacle, we consider collaborative clustering as an appropriate refinement of standard hierarchical clustering, where the joint output is sufficiently redacted, to allow for the necessary protection of parties’ privacy in the most meaningful manner, but minimally so, to preserve the gained utility. In particular, we introduce a security definition that is based on point-agnostic dendro- grams, which explicitly capture only the merging history and useful statistics of the formed clusters, thus ideally balanc- ing accuracy and privacy.

}

\Comment{
At the end of the protocol, the parties learn clusters' representatives and their sizes, while the input datasets are protected and not revealed during the cryptographic protocol.

We provide privacy definitions that minimize the amount of leakage of the input datasets, while including relevant information about the dendrogram computation.
}

}

\Comment{
\noindent\textbf{\underline{Attempt 3: Randomly permute nodes.}} We can further permute the nodes at each layer under a random permutation. This protects the mapping between inputs and clusters, but  we ``destroyed'' too much information. In particular, the structure of permuted dendrogram only shows that two clusters were merged in each iteration, but this is something that the participants know already! We thus need to preserve the dendrogram structure with more utility.


\ignore{
One way to remove the above leakage is by not only redacting  information from $T$, but also by changing the structure of $T$. In particular, after each layer $i$ is computed, its nodes are permuted under a randomly chosen permutation (without affecting the incoming edges from the previous layer). To highlight the usefulness of this, consider our simple inference attack from before. Due to the fact that the nodes at layer $n$ are permuted, no party can directly infer whether the merged clusters belong to the same party (or even whether they belong to themselves).}



\noindent\textbf{\underline{Attempt 4: Randomly permuting once.}} Finally, we propose to keep the dendrogram structure, but  permute nodes only once at the last layer. We believe that this represents the right balance between preserving the utility of the dendrogram, and reducing the amount of  revealed information. The parties can still infer in which order the clusters where merged  and the internal topology of the clusters, which has applications in medicine and biology~\cite{Eisen98,Girvan2002Community}. At the same time, this approach hides the mapping of points to clusters and the exact ordering of cluster merging, thus preventing simple attacks based on inferring distances between points.


\ignore{we are ready to explain our final approach (see Figure~\ref{f:dendrogram}~(right)) This is essentially the same as the previous one, but only the points at layer $n$ are permuted. On one hand, this avoids a simple inference attack but on the other hand it does not reveal information about subsequent layers in $T$. In particular, the parties can still infer in which order the cluster where merged  and the internal topology of the clusters -- one of the basic reasons to use hierarchical clustering.  We believe this instantiation of hierarchical clustering achieves the correct balance between utility (in order to motivate the use of hierarchical clustering) and privacy (in terms of limiting leakage).}

}

\section{Main Construction}\label{sec:main}

We now present our main construction,  protocol $\hcluster$ for {\bf P}rivate {\bf H}ieararchical {\bf C}lustering that securely realizes the ideal functionality $f^*_{HC}$ (of Figure~\ref{fig:idealhc}) when jointly run by parties $\Pone, \Ptwo$.

\Comment{
\smallskip\noindent {\bf General approach.} Designing efficient privacy-preserving protocols for complex learning tasks is technically challenging, and hierarchical clustering---an inherently iterative process with cubic costs in its input size---bears no exception. Carrying out such a laborious task in its entirety by computing over ciphertext (malleable in some operation-complete encryption form, e.g., Yao's GC or FHE), would render our solution largely impractical, if not infeasible.}

\smallskip\noindent {\bf General approach.} As discussed earlier, for efficiency reasons, we seek to avoid carrying out hierarchical clustering---a complex and inherently iterative process of cubic costs---in its entirety by computing over ciphertext (e.g., via GC or FHE). Instead, we adopt a \emph{mixed-protocols} design, 
decomposing hierarchical clustering into more elementary tasks. We then use tailored secure and efficient protocols for each task, and combine these components into a final protocol, in ways that minimize the cost in converting data encoding between individual sub-protocols. Hence, our solution is a secure mixed-protocol specifically tailored for hierarchical clustering.

It is worth noting that generic solutions from 2-party computation (2PC) (e.g.,~\cite{ABY}), would solve the problem but would not easily scale to large datasets. During hierarchical clustering, we need to maintain a distance matrix between two parties with space complexity $O(n^2)$.
If one relies solely on a single generic approach such as GC or secret sharing, the communication bandwidth would become the bottleneck. Hence, using additively homomorphic encryption during our protocol's setup phase in order to produce a ``shared permuted'' distance matrix allows us not only to hide the correspondence between euclidean distances and original points, but also to be more communication efficient eventually.
Another advantage compared to other 2PC techniques is that our approach can achieve better precision as we explain in more detail in Section~\ref{sec:eval}.

Our protocol securely implements $f^*_{HC}$ for the configuration that the parties specify: linkage $\delta(\cdot,\cdot)$, termination condition $\mathsf{End}(\cdot,t)$, cluster statistics set $M$, selection function~$\mathsf{S(\cdot)}$. Yet for simplicity, hereby, we use the following \emph{default configurations}, where:
\begin{enumerate}
    \item \emph{complete linkage} over \emph{one-dimensional} data is used;
    \item the termination condition results in \emph{$t$ target nodes}; 
    \item \emph{target selection} is used for safe metadata selection; and
    \item \emph{only} representative values and size statistics are learned.
\end{enumerate}
That is, by (2) - (4) in what follows (and in our experiments  in Section~\ref{sec:eval}), the set of redacted statistics $M^*$ consists  of the representatives $rep_1,\ldots,rep_{\ell_t}$ and sizes $size_1,\ldots,size_{\ell_t}$ of $\ell_t=t$ target clusters \textcolor{black}{(recall that representatives are a predefined type of centroid of the cluster, e.g., average or median)}, where $t$ is fixed in advance. Configuration 1) is used only for clarity; we discuss optimizations for single linkage and extensions to higher dimensions in Section~\ref{sec:analysis} (and we report on the evaluation of such extensions in Section~\ref{sec:eval}). 

\begin{algorithm}[t!]
	\caption{$\hcluster$: Private Cluster Analysis}\label{alg:protocol}
	\begin{footnotesize}
          \PoneInput{$P=\{p_1, \dots, p_{n_1}\}$, security
                  parameter $\lambda$}
		\PtwoInput{$Q=\{q_1, \dots, q_{n_2}\}$, security
                  parameter $\lambda$}
                \COutput{Merging history, $\{rep(\cdot)$, $size(\cdot)\}$ of $t$ target nodes}       
\CPara{Default configurations}
		\linespace
                $\Pone$: Generate $(\pk, \sk) \larr \gen(1^\lambda)$;
                send $\pk$ to $\Ptwo$

                $\Ptwo$: Generate $(\pk', \sk') \larr \gen(1^\lambda)$;
                send $\pk'$ to $\Pone$

                $\Pone, \Ptwo$: Jointly run $\hcluster.\setup$,
                $\hcluster.\cluster$, $\hcluster.\filter$


	\end{footnotesize}
      \end{algorithm}


\smallskip\noindent {\bf Protocol overview.} After choosing configurations, $\Pone, \Ptwo$ run protocol $\hcluster$ (Algorithm~\ref{alg:protocol}),
 with inputs their datasets $P, Q$ of $n_1, n_2$ points, $n=n_1 + n_2$, security parameter $\lambda$,  
and statistical parameter $\kappa$.
Each party establishes its individual Paillier key-pair, and then parties exchange their corresponding public keys. 

\begin{figure}[t]
		\centering
		\includegraphics[angle=0, width = 0.46\textwidth]{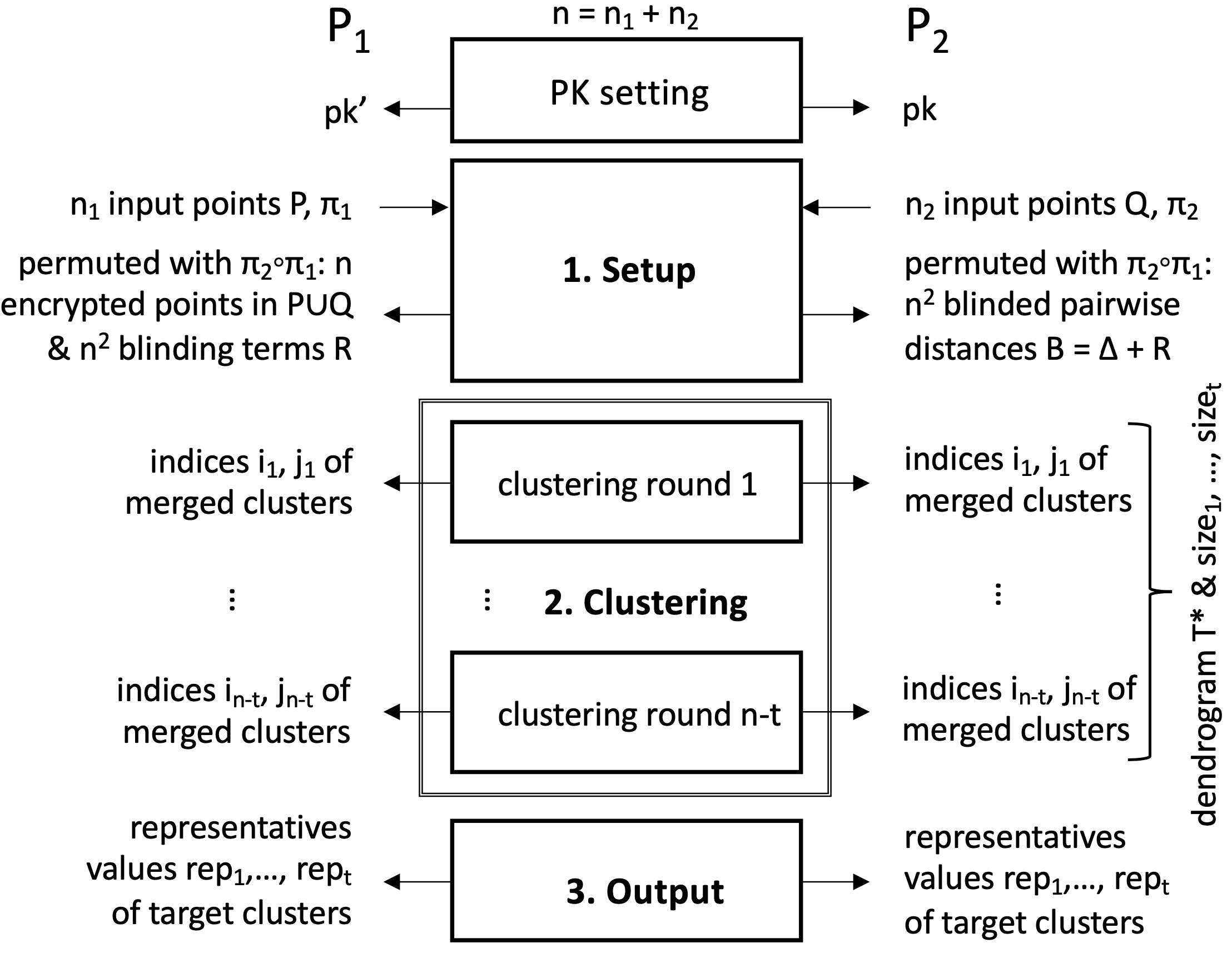}
	\vspace{-.4cm}
	\caption{Overall workflow of our protocol $\hcluster$.}
	\label{fig:flow}
\end{figure}

Then, parties run sub-protocols $\hcluster.\setup$, $\hcluster.\cluster$ and $\hcluster.\filter$, which comprise the three main phases in our protocol, in direct analogy to the three components of~$f^*_{HC}$. The general flow of our protocol is described below, in reference to also Figure~\ref{fig:flow}.

In a setup phase, sub-protocol $\hcluster.\setup$ processes the $n$ input points, viewed as an input array $I$, and all pairwise distances among these points, viewed as a $n\times n$ cluster distance matrix $\Delta$. Here, $I, \Delta$ are only virtual, corresponding to an early joint state of $\Pone, \Ptwo$ that is actually \emph{secret-shared} between them. Specifically, $\Pone$ holds an array $L$ with exactly $I$'s elements but each \emph{AHE-encrypted} under $\Ptwo$'s secret key, and a $n\times n$ matrix $R$ with \emph{random blinding terms}, whereas $\Ptwo$ holds 
the matrix $B = \Delta + R$ with \emph{blinded pairwise cluster distances}. Importantly, as $f^*_{HC}$ specifies, the joint state $\{I, \Delta\}$ is split only after $I$'s elements and $\Delta$'s rows and columns are \emph{randomly shuffled}, with $\Pone, \Ptwo$ not knowing the exact shuffling used.



In a clustering phase, sub-protocol $\hcluster.\cluster$ virtually runs the ordinary hierarchical clustering algorithm $\mathsf{HCAlg}$ on matrix~$\Delta$: $\Pone, \Ptwo$ process their individual states $R, B$ to iteratively merge singletons into target clusters, based on inter-cluster distances in $\Delta$. Each iteration merges two clusters into a new one via three tasks:

\begin{itemize}[leftmargin=10pt]
\item {\bf Find pair:} First, $\Pone, \Ptwo$ find the closest-cluster pair $(i,j)=\mathsf{argMin}(\Delta)$, $i<j$, to merge, i.e., the indices in $\Delta$ of the minimum inter-cluster distance~$D_{ij}$.

\item {\bf Update linkages:} Then, $\Pone, \Ptwo$ update $\Delta$ to $\Delta'=B'-R'$ with the new cluster distances after pair $(i,j)$ is merged into cluster $C=C_i\cup C_j$. This entails computing (and splitting via a fresh blinding term) distance $\delta(C,C')$ between $C$ and each not-merged cluster $C'$, which equals to the largest.smallest) of $\delta(C_i,C')$ and $\delta(C_j,C')$ (by associativity of the $\mathsf{max}$/$\mathsf{min}$ operator).

\item {\bf Record merging:} Finally, $\Pone, \Ptwo$ record in $\Delta '$ that the new cluster $C$ is formed by merging $C_i$ and~$C_j$.

\end{itemize}

In an output phase, sub-protocol $\hcluster.\filter$ processes the final  state $\{I, \Delta\}$ to compute the merging history and metadata for all safe (target) clusters. As Figure~\ref{fig:flow} indicates, conceptually the  output can be considered to be computed in two phases: During clustering, the indices of merged clusters learned after each clustering round collectively encode information about the dendrogram $T^*$ and the sizes of the $t$ target clusters. The output phase solely computes the representative values of these clusters. This view is accurate enough  to ease presentation but, as we discuss later, the exact details involve processing of carefully recorded data, after each one of the $n-t$ cluster-merging rounds executed during the clustering phase.

A main consideration when devising our protocol was to improve efficiency via a modular design, where separate parts can be securely achieved via different techniques.
By securely splitting the joint state $\{I, \Delta\}$ into $\{L, R\}$, $B$, we can implement all protocol components that involve (distance or metadata) computations over points using Paillier-based AHE, except when computing $\mathsf{max}$ (or $\mathsf{min}$), for which we rely on GC. Conveniently, all protocol components required by the setup phase to form the joint state $\{I, \Delta\}$, namely to construct, shuffle and split $\{I, \Delta\}$ into $\{L, R\}$, $B$, can be securely implemented by relying on homomorphic encryption. 

We next provide more details on how each component is implemented. We assume points are unambiguously mapped into integers in $\Z_N$ and all homomorphic (resp. plaintext) operations are reduced modulo $N^2$ (resp. $N$). We consistently denote the AHE-encrypted, under $\pk$ (resp. $\pk'$), plaintext $d$ by $\enOne{d}$ (resp. $\enTwo{d}$) and the AHE-decrypted, under any key, ciphertext $c$ by $\de{c}$. Whenever the context is clear, we  denote each of the two $n\times n$ matrices $R$, $B$ (maintained by $\Pone, \Ptwo$) by $\Sigma$. Finally, we denote the joint execution by $\Pone, \Ptwo$ of a GC-based protocol $\mathsf{GC}$, on private inputs $I_1, I_2$ to get private outputs $O_1,O_2$, by $(O_1;O_2)\leftarrow \mathsf{GC}(I_1;I_2)$. Figure~\ref{fig:notation} summarizes the used notation by our detailed protocol descriptions.

\begin{figure}
\centering
\footnotesize
\begin{tabular}{|c | l|}
	\hline
$\lambda, \kappa$ & security and statistical parameters\\
$t, \ell_t$ & termination parameter, \# of target clusters\\
$(\pk, \sk), (\pk', \sk')$  & public and secret keys of parties $\Pone, \Ptwo$\\
$\enOne{d}, \enTwo{d}$ & AHE-encrypted plaintext $d$ under $\pk, \pk'$ \\
$\de{c}$ & AHE-decrypted ciphertext $c$\\
$\Sigma$ & $n\times n$ matrices $R$, $B$ stored by $\Pone, \Ptwo$\\
$(O_1;O_2)\leftarrow \mathsf{GC}(I_1;I_2)$ & 
$\Pone, \Ptwo$ run $\mathsf{GC}$ on $I_1, I_2$ to get $O_1,O_2$\\
$\pi_1,\pi_2$ & permutations contributed by $\Pone, \Ptwo$\\
$\dist(p, q)$ & square Euclidean distance of $p, q$\\
$rep(\cdot), size(\cdot)$ & representatives and sizes of clusters\\
\hline
\end{tabular}
\vspace{-.4cm}
\caption{Basic notation used in our protocol $\hcluster$.}\label{fig:notation}
\end{figure}

\smallskip\noindent {\bf Setup phase.} $\Pone, \Ptwo$ set up their states in three rounds of interactions, as shown in Algorithm~\ref{alg:cluster_setup}, using only homomorphic operations over AHE-encrypted data and contributing equally to the randomized state permutation and splitting.
Initially, $\Ptwo$ prepares, encrypts under its own key and sends to $\Pone$, information related to its input set $Q$, which includes its encrypted points among other helper information $H$, and their encrypted pairwise distances $D$ (lines 1-4).

Then, $\Pone$ is tasked to initialize the states $L$, $R$ and $B$. First, the list $L$ of all encrypted (under $\pk'$) points in $P\cup Q$ is created (by arranging the sets in some fixed ordering and then concatenating $Q$ after $P$), and all points are further blinded by random additive terms in $S$ (lines 5-9). Similarly, the matrix $B$ of encrypted (also under $\pk'$) pairwise distances is computed (using the ordering induced by $L$ to arrange the points), and all distances are blinded by random additive terms in $S$ (lines 10-14). The computation of square Euclidean distances across sets $P, Q$ (line 12, using also elements in $H$) and the blinding of $L$ and $B$ (lines 8, 13) are all performed in the ciphertext domain via the homomorphic property of AHE encryption. All blinding terms in $S$ and $R$ are then encrypted (each under $pk$, lines 9, 14) and $S$, $L$, $R$ and $B$ are sent to $\Ptwo$, after their elements are shuffled using a random permutation $\pi_1$ (line~15).

Finally, $\Ptwo$ roughly mirrors this by further blinding the encrypted points in $L$ and $\Pone$'s encrypted terms in $S$ by random additive terms in $S'$ (both in the ciphertext domain, lines 16-19) and also blinding the encrypted distances in $B$ and $\Pone$'s encrypted terms in $R$ by random additive terms in $R'$ (the former in the plaintext domain and the latter in the ciphertext domain, lines 20-22). The freshly blinded $S$, $L$, $R$ are sent to $\Pone$, after their elements are shuffled using a random permutation $\pi_2$ (line~23). Finally, $\Pone$ decrypts the mutually-contributed blinding terms in $S$ and $R$, and uses the recovered values in $S$ to completely remove the  terms from $L$ (in the ciphertext domain, by the properties of AHE encryption, lines 24-27). Due to this, permutation $\pi_2 \circ \pi_1$ looks completely random to both parties, while they have securely split joint state $\{I, \Delta\}$ into $\{L, R\}$, $B$.

\begin{algorithm}[t!]
	\caption{$\hcluster.\setup$: Setup Phase}\label{alg:cluster_setup}
	\begin{footnotesize}
        
        
        
		$\Ptwo$: \hspace*{\fill}\%\texttt{Create \& send helper info}

            \textcolor{black}{    Compute matrix $\vH$: $H_{1,i}=
                \enTwo{q_i}$, $H_{2,i} = \enTwo{-2q_i}$, $H_{3,i} =
                \enTwo{q_i^2}$ \hspace*{\fill}$i \in [1:n_2]$}

	\textcolor{black}{	Compute matrix $\vD$: $D_{i,j} =
                \enTwo{\dist(q_i, q_{j})}$ \hspace*{\fill}$i,j \in [1:n_2]$}

              	Send $\{{\vH}, \vD\}$ to $\Pone$ 

		$\Pone$:  \hspace*{\fill}\%\texttt{Blind
                  points, linkages}

                Compute array $\vS$: $S_i = s_i$,
                $s_i \xleftarrow{\$}  \{0,1\}^\kappa$ \hspace*{\fill}$i \in [1:n]$

                Compute array $\vL$: $L_i = \enTwo{p_i}$,  if $i \le
                n_1$;  else $L_i = H_{1,i-{n_1}}$ 

                Blind $\vL$ as: $L_i:= L_i\cdot
                \enTwo{S_i}$ 

                Encrypt $\vS$ as:
                $S_i := \enOne{S_i} $ 

                 Compute matrix $\vR$: $R_{i,j} = r_{i,j}$,
                 $r_{i,j}\xleftarrow{\$}  \{0,1\}^\kappa$ \hspace*{\fill}$i,j \in
                 [1:n]$

                 Compute matrix $\vB$: $B_{i,j} = \enTwo{\dist(p_i, p_{j})}$, if $i,j \le
                n_1$;

               \textcolor{black}{ $B_{i,j} = D_{i-n_1,j-n_1} $, if
                $n_1< i,j$; else for $i<j$, $B_{i,j} = \enTwo{p^2_i}\cdot H_{2,j}^{p_i}\cdot H_{3,j}$ }

                Blind $\vB$ as:
                $B_{ij}:= B_{ij}\cdot \enTwo{R_{i,j}}$ 

                Encrypt $\vR$ as: $R_{i,j} := \enOne{R_{i,j}} $

                Permute $\vS$, $\vL$, $\vR$ and $\vB$ via a random permutation
                $\pi_1(n)$

                Send $\{\vS, \vL, \vR, \vB\}$ to
                $\Ptwo$ \hspace*{\fill}\%\texttt{Send permuted blinded
                  data}
		
		$\Ptwo$: \hspace*{\fill}\%\texttt{Blind
                  received data}

                Compute array $\vS'$: $S'_i = s'_i$,
                $s'_i \xleftarrow{\$}  \{0,1\}^\kappa$ \hspace*{\fill} $i \in [1:n]$

                Blind $\vL$ as: $L_i:= L_i\cdot
                \enTwo{S'_i}$ 

                Blind $\vS$ as: $S_i:= S_i\cdot
                \enOne{S'_i}$ 

                Compute matrix $\vR'$: $R'_{i,j} = r'_{i,j}$,
                $r'_{i,j}\xleftarrow{\$}  \{0,1\}^\kappa$ \hspace*{\fill}$i,j \in [1:n]$

                Decrypt and re-blind matrix $\vB$: $B_{i,j} =
                \de{B_{i,j}} + R'_{i,j}$ \label{line:decrypt1}

                Blind $\vR$ as:
                $R_{i,j} := R_{i,j} \cdot \enOne{R'_{i,j}} $

                Permute $\vS$, $\vL$ and $\vR$ via a random permutation
                $\pi_2(n)$

                Send $\{\vS, \vL, \vR\}$ to $\Pone$
                \hspace*{\fill}\%\texttt{Send permuted points and blinding terms}
                
		$\Pone$: \hspace*{\fill}\%\texttt{Store permuted points \&
                  linkages' blinding terms}

                Decrypt $\vS$ as:
                $S_i := \de{S_i} $ \hspace*{\fill}$i\in [1:n]$

                Unblind $\vL$ as:
                $L_i := L_i \cdot \enTwo{S_i}^{-1}$
                
                Decrypt $\vR$ as:
                $R_{i,j} := \de{R_{i,j}} $ \hspace*{\fill}$i,j\in [1:n]$\label{line:decrypt2}
       
	\end{footnotesize}
      \end{algorithm}


\begin{algorithm}[t]
\caption{$\hcluster.\cluster$: Clustering Phase}\label{alg:cluster_cluster}
 
	\begin{footnotesize}
	
          $\Pone$, $\Ptwo$:  

          Initialize merging history: $\Sigma_{i,i} = (i,\perp)$\hspace*{\fill}$i \in [1:n]$
         
          Initialize: $\ell = 1$, $\ell_t = t$
		
          \Repeat{$\ell > n-\ell_t $}{
            Jointly run 
            $(i,j;i,j)\leftarrow \argMinSelect(\vR;\vB)$, $i<j$\hspace*{\fill}\%\texttt{Find pair}


			\ForEach{$k =1,\dots,n$, $k \ne i, j$}{
				\If{ $\Sigma_{i,k}\neq \perp$ \emph{and} $\Sigma_{j,k}\neq \perp$ } {

                                  $\Pone$:  $X \xleftarrow{\$} \{0,1\}^\kappa$
                                  \hspace*{\fill}\%\texttt{Pick new
                                    blinding term}
                                  

                                   $\Pone$, $\Ptwo$: Jointly run \hspace*{\fill}\%\texttt{Find re-blinded
              max linkage}
                                  \hspace*{\fill}$(\perp;Y)\leftarrow\maxDist(R_{i,k},
                                  R_{j,k}, X; B_{i,k}, B_{j,k})$

                                  $\Pone$: Set: $R_{ik}=X$, $R_{ki}=X$ \hspace*{\fill}\%\texttt{Update linkages}

                                  $\Ptwo$: Set: $B_{ik}=Y$, $B_{ki}=Y$

                                }
                    }

          Set: $\Sigma_{j,j} := ((\Sigma_{j,j},\ell),i)$,
          $\Sigma_{i,i} := ((\Sigma_{i,i}, \Sigma_{j,j},
          \ell),\perp)$

         Set: $\Sigma_{k,j}=\perp$, $\Sigma_{j,k}=\perp$
          \hspace*{\fill} $k\in [1:n]\setminus\{j\}$

           Set: $\ell := \ell+1$ \hspace*{\fill}\%\texttt{Record
            merging}
		}

	\end{footnotesize}

\end{algorithm}

\medskip\noindent {\bf Clustering phase.} Once $\Pone, \Ptwo$ have set up their states, they run the hierarchical clustering iterative process ( Algorithm~\ref{alg:cluster_cluster}) operating solely on their individual matrices $R$, $B$ via two special-purpose GC-based protocols for secure comparison.
Importantly, each party encodes cluster information in the diagonal of its matrix state $\Sigma$; initially, the $i$-th entry stores $(i, \perp)$, denoting the (never-merged but already permuted) singleton of rank $i$.
Hierarchical clustering runs in exactly $n-\ell_t=n-t$ iterations, or clustering rounds.

First, at the start of each iteration, $\Pone, \Ptwo$ find which pair of clusters must be merged by jointly running the GC-protocol $(i,j;i,j)\leftarrow \argMinSelect(R;B)$ (line 5): The parties contribute their individual matrices $R, B$ of blinding terms and blinded linkages, to learn the indices $(i,j)$ of the minimum value $B_{i,j}-R_{i,j}$, with $i<j$ by convention (since $R, B$ are symmetric matrices). The garbled circuit for $\argMinSelect$ first removes the blinding terms by computing $D=B-R$, compares all values in $D$ to find the minimum element $D_{i,j}=\min_{x,y} B_{x,y}$, and returns to both parties the indices $i,j$.
%
Next, once pair $(i,j)$ is known to $\Pone, \Ptwo$, they proceed to jointly update the linkages (lines 7-12). For each cluster $k$ in $\Sigma$, they change its linkage to the newly merged cluster as the maximum between its linkages to clusters $i$, $j$, by jointly running the GC-protocol $(\perp;Y)\leftarrow\maxDist(R_{i,k}, R_{j,k}, X; B_{i,k}, B_{j,k})$ (line 10): The parties contribute the two entries from their individual matrices $R, B$ that are needed for comparing the linkages $B_{i,k}-R_{i,k}$, $ B_{j,k}-R_{j,k}$ between cluster $k$ and clusters $i$, $j$, and $\Ptwo$ learns the maximum value of the two but blinded by the random blinding term $X$ inputted by $\Pone$. The garbled circuit for $\maxDist$ simply returns to (only) $\Ptwo$ the value $\max\{B_{i,k}-R_{i,k}, B_{j,k}-R_{j,k}\}+X$.
Finally, at the end of iteration $\ell$, $\Pone, \Ptwo$ record information about the merging of clusters $i$, $j$, $i<j$ (lines 14-15): By convention, the new cluster is stored at location $i$, by adding the rank $\ell$ and the information stored at location $j$ (updated with a pointer to $i$), and by deleting all distances related to cluster $j$. Overall, the full merging history is recorded.

\iffull In Appendix~\ref{app:sub}, we provide details on our  implementation of GC-protocols $\argMinSelect$, $\maxDist$, also used in~\cite{min1,min2,popa,privsim}.

\fi

\smallskip\noindent {\bf Output phase.} Once clustering is over, $\Pone$, $\Ptwo$ compute in two rounds of interaction  (Algorithm~\ref{alg:cluster_filter}) the common output, consisting of the merging history and the representatives and sizes of the target clusters 
using homomorphic operations over encrypted data.

\begin{algorithm}[t]
\caption{$\hcluster.\filter$: Output Phase}\label{alg:cluster_filter}
	\begin{footnotesize}
	
          $\Pone$: \hspace*{\fill}\%\texttt{Compute encrypted point averages}

          Initialize arrays $\vE$, $\vJ$: $E_i=J_i = \perp$ \hspace*{\fill}$i \in [1:n]$
		
		\ForEach{$i=1,\dots,n$}{
			\If{$R_{i,i}$ \emph{encodes a target
                            cluster $C_i$}}{
                          Find the index set $I_i$ of points in
                          cluster $C_i$

                         Set $E_i = \prod_{j\in I_i}L_j$, $J_i = |I_i|$
			}
		}
		Send $\{\vE, \vJ\}$ to $\Ptwo$

		$\Ptwo$: \hspace*{\fill}\%\texttt{Compute point averages}

                Decrypt $\vE$ as: $E_i := \langle E_i \rangle$
                \hspace*{\fill}$i \in [1:n]$

                Send $\vE$ to $\Pone$
                
                $\Pone$, $\Ptwo$: \hspace*{\fill}\%\texttt{Return output}

                Output $\{\Sigma_{i,i}, E_i/|J_i|,
                |J_i|\}$  \hspace*{\fill}$i \in [1:n]$

	\end{footnotesize}
\end{algorithm}


First, $\Pone$ computes encrypted point averages in all target clusters, by exploiting the homomorphic properties of AHE  (lines 1-7): Using the diagonal in matrix $R$, $\Pone$ first identifies each (of $t$ total) target cluster $C_i$ and then finds the index set $I_i$ (over permuted input points $\pi_2 \circ \pi_1(P\cup Q)$) of the points contained in $C_i$, to finally compute $\prod_{j\in I_i}L_j =\prod_{j\in I_i}\enTwo{p_j}$. The resulted $t$ encrypted point averages and cluster sizes are sent to $\Ptwo$, who returns to $\Pone$ the $t$ plaintext point averages, i.e., $\sum_{j\in I_i} p_j=\sum_{p_j\in C_i} p_j$ for each $C_i$ (lines 8-10). At this point, both parties can form the common output (line 12).

\Comment{

The following theorem captures our protocol security.

\begin{theorem}\label{THM:MAIN}
  Assuming Paillier's encryption scheme is semantically secure and that $\argMinSelect$ and $\maxDist$ are securely realized by GC-based protocols, protocol $\hcluster$ securely realizes functionality $f^*_{HC}$ as per Definition~\ref{def:sec}.  \end{theorem}

\smallskip\noindent {\bf Asymptotic complexity.} Assuming that each cryptographic operation takes $O(1)$ time (i.e., ignoring its dependence on the security parameter $\lambda$), the asymptotic performance costs incurred on $\Pone, \Ptwo$, during execution of $\hcluster$, are as follows. In setup phase, the dominant cost for both parties comes from performing $O(n^2)$ cryptographic operations (in order to each populate its individual state $\Sigma$). In clustering phase, the cost of each of the total $\ell_t=O(n)$ iterations is dominated by the complexity of GC-based protocols $\argMinSelect$, $\maxDist$, where the cost of garbling and evaluating a circuit $C$, with a total number of wires $|C|$, is $O(|C|)$. Thus, during the $\ell$-th iteration: evaluating circuit $\argMinSelect$ entails $n^2-2\ell$ comparisons of $d^2$-bit values (of cluster distances) and subtractions of $\lambda$-bit values (of blinding terms), for a total size of $O(\lambda(n^2-\ell))$, and likewise, evaluating
circuit  $\maxDist$ entails comparisons of $\lambda$-bit values, and $n^2-2\ell$ such circuits are evaluated at round $\ell$. In output phase, the cost for both parties is $O(\ell_t)=O(n)$. Thus, asymptotically, the total performance cost for both parties is $O(\lambda n^3)$, i.e., the same as the cost of ordinary hierarchical clustering multiplied by a factor $\lambda$ due to the use of cryptography.

\smallskip\noindent {\bf Single linkage.}
We discuss how our protocol can be optimized
if we focus only on single linkage. In particular, we present a modified version of the protocol  that runs in time $O(\kappa n^2)$, as opposed to $O(\kappa n^3)$.


Our main protocol performs a quadratic number of comparisons in every round of clustering to identify the closest clusters. This takes place with the sub-protocol $\mathsf{ArgminSelect}$ (steps 6-7 in Algorithm~\ref{alg:cluster}) which compares all the values in matrix $\vV$. This ``naive'' approach works independently of the linkage function chosen. However, it turns out that for the case of single linkage we can use a much faster alternative that reduces the number of comparisons per round to $O(n)$ (as opposed to $O(n^2)$). This technique is well-known (e.g., see~\cite[Section 17.2.1]{DBLP:books/daglib/0021593}) and we explain it briefly.

After computing the encrypted distance matrix $\vV$, the two parties compute the minimum distance per matrix row and store it in a separate array $\rowmin$ ($\rowmin[i]$ stores the minimum value in the $i$-th row of $\vV$). We note that this can be achieved by essentially running the $\mathsf{ArgminSelect}$ protocol separately per matrix row.
Then, during every clustering round \emph{instead} of running the $\mathsf{ArgminSelect}$ protocol (steps 6-7 in Algorithm~\ref{alg:cluster}) in order to locate the minimum value in $\vV$, the two parties proceed as follows:
\begin{enumerate}[noitemsep,topsep=0pt,parsep=0pt,partopsep=0pt]
	\item $\Pone,\Ptwo$ run $\mathsf{ArgminSelect}$ over the values of $\rowmin$ and get the index $i$ of the cluster that corresponds to the row of $\vV$ that contains the minimum values.
	\item $\Pone,\Ptwo$ run $\mathsf{ArgminSelect}$ over the values from the $i$-th row of $\vV$ and get the index $j$ of the closest cluster to cluster $i$.
\end{enumerate}

Moreover, after updating the inter-cluster distances in $\vV$ (steps 10-17 in Algorithm~\ref{alg:cluster}) the two parties have to update array $\rowmin$. This is done as follows:
\begin{enumerate}[noitemsep,topsep=0pt,parsep=0pt,partopsep=0pt]
	\item
	$\Pone,\Ptwo$ repeatedly run $\minselect$ over the $i$-th row of $\vV$  and store the minimum distance in $\rowmin[i]$.

	\item $\Ptwo$ sets $\rowmin[j]$ to $\bot$.
	\item For every  $k\neq i$, $\Pone,\Ptwo$ run $\minselect$ comparing $\rowmin[k]$ with $V_{ki}$  and store the result in $\rowmin[k]$.

\end{enumerate}
Note that the above steps require at most $4(n-1)$ comparisons, as opposed to at most $n^2/2-1$  needed by our basic protocol to identify the minimum distance index. In practice, this results in significant improvements to the performance of our privacy-preserving hierarchical clustering solution, as demonstrated in Section~\ref{sec:eval}, since all these comparisons are  done using garbled circuits.


%

\smallskip\noindent {\bf High-dimensional data.}
So far we considered single-dimensional data ($d=1$). In practice, hierarchical clustering is mostly applied to higher-dimension data ($d > 1$) and our protocol can be easily adapted. The core part of our protocol deals with comparisons between squared Euclidean distances, therefore it remains entirely unaffected by the number of dimensions. The modifications thus have to do with the setup process and the representative computation.


Initially, $\Ptwo$ represents each point not by three but by $3d$ encryptions, essentially running step 4 of Algorithm 1 independently for each dimension. Thus, list $\vQ$ consists of $dn_2$ encryptions and $\vQ'$ of $2dn_2$. Then, $\Pone$ does the same for his points and computes the values $M_{ij}$ (step 15) as the sum of the per dimension computation (which can be achieved with AHE). The shuffling process remains largely unaffected, other than the fact that lists $\vL,\vS,\vS'$ consist of $dn$ encryptions instead of $n$. Finally, the representative computation (Algorithm~\ref{alg:cluster}, step 23) needs to compute $E_i$ as a vector of $d$ values.





\smallskip\noindent {\bf Generalizations.} Our protocol can be extended to other distance metrics, including $L_1$, $L_2$ or Euclidian, and in general any $L_p$ distance for $p \ge 1$. The only modification is computing the pairwise distance matrix during setup. Whereas our chosen squared Euclidian distance metric enables joint computation of pairwise distances  solely using AHE and with a simple interaction between among $\Pone, \Ptwo$, if other distance metrics are considered, then the design of more elaborate protocol components may be required in setup phase.

Extensions towards maintaining security also in the malicious threat model, where parties may misbehave arbitrarily, are generally possible---though, with new design challenges emerged. Techniques that can be applied include augmenting homomorphic ciphertexts with zero-knowledge proofs and garbled circuits with cut-and-choose extensions, as well as, recent developments in efficient secure computation~\cite{DBLP:conf/ccs/WangRK17}.

Extending our protocol to support multiple participants, as typically considered by federated learning, is our main direction for future work. We believe that this goal is feasible but would require drastically different techniques to achieve practical performance. Techniques that may help in this direction include replacing the garbled circuit components with secret-sharing based protocols~\cite{Araki16} and using recent advancements in secure computation~\cite{global,overdrive}.


\subsection{Scaling to multiple dimensions}\label{sec:multi-d}

So far we considered single-dimensional data ($d=1$). In practice, hierarchical clustering is mostly applied to higher-dimension data ($d > 1$) and our protocol can be easily adapted. The core part of our protocol deals with comparisons between squared Euclidean distances, therefore it remains entirely unaffected by the number of dimensions. The modifications thus have to do with the setup process and the representative computation.


Initially, $\Ptwo$ represents each point not by three but by $3d$ encryptions, essentially running step 4 of Algorithm 1 independently for each dimension. Thus, list $\vQ$ consists of $dn_2$ encryptions and $\vQ'$ of $2dn_2$. Then, $\Pone$ does the same for his points and computes the values $M_{ij}$ (step 15) as the sum of the per dimension computation (which can be achieved with AHE). The shuffling process remains largely unaffected, other than the fact that lists $\vL,\vS,\vS'$ consist of $dn$ encryptions instead of $n$. Finally, the representative computation (Algorithm~\ref{alg:cluster}, step 23) needs to compute $E_i$ as a vector of $d$ values.





\subsection{Optimization for single linkage}\label{sec:optimized}
We discuss how our protocol can be optimized
if we focus only on single linkage. In particular, we present a modified version of the protocol  that runs in time $O(\kappa n^2)$, as opposed to $O(\kappa n^3)$.


Our main protocol performs a quadratic number of comparisons in every round of clustering to identify the closest clusters. This takes place with the sub-protocol $\mathsf{ArgminSelect}$ (steps 6-7 in Algorithm~\ref{alg:cluster}) which compares all the values in matrix $\vV$. This ``naive'' approach works independently of the linkage function chosen. However, it turns out that for the case of single linkage we can use a much faster alternative that reduces the number of comparisons per round to $O(n)$ (as opposed to $O(n^2)$). This technique is well-known (e.g., see~\cite[Section 17.2.1]{DBLP:books/daglib/0021593}) and we explain it briefly.

After computing the encrypted distance matrix $\vV$, the two parties compute the minimum distance per matrix row and store it in a separate array $\rowmin$ ($\rowmin[i]$ stores the minimum value in the $i$-th row of $\vV$). We note that this can be achieved by essentially running the $\mathsf{ArgminSelect}$ protocol separately per matrix row.
Then, during every clustering round \emph{instead} of running the $\mathsf{ArgminSelect}$ protocol (steps 6-7 in Algorithm~\ref{alg:cluster}) in order to locate the minimum value in $\vV$, the two parties proceed as follows:
\begin{enumerate}[noitemsep,topsep=0pt,parsep=0pt,partopsep=0pt]
	\item $\Pone,\Ptwo$ run $\mathsf{ArgminSelect}$ over the values of $\rowmin$ and get the index $i$ of the cluster that corresponds to the row of $\vV$ that contains the minimum values.
	\item $\Pone,\Ptwo$ run $\mathsf{ArgminSelect}$ over the values from the $i$-th row of $\vV$ and get the index $j$ of the closest cluster to cluster $i$.
\end{enumerate}

Moreover, after updating the inter-cluster distances in $\vV$ (steps 10-17 in Algorithm~\ref{alg:cluster}) the two parties have to update array $\rowmin$. This is done as follows:
\begin{enumerate}[noitemsep,topsep=0pt,parsep=0pt,partopsep=0pt]
	\item
	$\Pone,\Ptwo$ repeatedly run $\minselect$ over the $i$-th row of $\vV$  and store the minimum distance in $\rowmin[i]$.

	\item $\Ptwo$ sets $\rowmin[j]$ to $\bot$.
	\item For every  $k\neq i$, $\Pone,\Ptwo$ run $\minselect$ comparing $\rowmin[k]$ with $V_{ki}$  and store the result in $\rowmin[k]$.

\end{enumerate}
Note that the above steps require at most $4(n-1)$ comparisons, as opposed to at most $n^2/2-1$  needed by our basic protocol to identify the minimum distance index. In practice, this results in significant improvements to the performance of our privacy-preserving hierarchical clustering solution, as demonstrated in Section~\ref{sec:eval}, since all these comparisons are  done using garbled circuits.


%
}


\section{Protocol Analysis}\label{sec:analysis}
\noindent {\bf Efficiency.} Asymptotically, our protocol achieves optimal performance, as it incurs no
extra overheads to the performance costs associated with running
HC (ignoring the
dependency on the security parameter $\lambda$).
The asymptotic
overheads incurred on $\Pone, \Ptwo$, during execution of each phase
of $\hcluster$, are as follows: In setup phase, the cost overhead for
each party is $O(n^2)$, primarily related to the cryptographic
operations needed to populate its individual state $\Sigma$. In
clustering phase, each of the $n-\ell_t=O(n)$ total iterations incurs
costs proportional the complexity of running GC-based protocols
$\argMinSelect$, $\maxDist$, where the cost of garbling and evaluating
a circuit $C$, with a total number of wires $|C|$, is $O(|C|)$. Thus,
during the $\ell$-th iteration: Evaluating circuit $\argMinSelect$
entails $n^2-2\ell$ comparisons of $l^2$-bit values (of cluster
distances) and subtractions of $\kappa$-bit values (of blinding
terms), for a total size of $O(\kappa(n^2-\ell))$; likewise,
evaluating circuit $\maxDist$ entails a constant number of comparisons
of $\kappa$-bit values and $O(n)$ such circuits are evaluated at iteration
$\ell$; thus, the total cost during this phase is $O(\kappa n^3)$ for
each party. In output phase, the cost is $O(\ell_t)=O(n)$ for each
party. Thus, the total running time for both parties is
$O(\kappa n^3)$.
\textcolor{black}{
Communication consists of $O(n^2)$ ciphertexts during setup (encrypted distances), $O(\kappa n^2)$ during each clustering round (for the garbled circuits' truth tables) and $O(n^2)$ ciphertexts during the output phase.}

\smallskip\noindent {\bf Optimized single-linkage protocol \opt.} As
described, 
our protocol exploits the associativity of operator $\mathsf{max}$ to
update the complete linkage between newly formed clusters $C$ and
other clusters $C'$, as the $\mathsf{max}$ of the linkages between
$C$'s constituent clusters and $C'$, securely realized via GC-protocol
$(\cdot;\cdot)\leftarrow\maxDist(\cdot; \cdot)$. Single linkages can
be supported readily by updating inter-cluster distances between $C$
and $C'$ as the $\mathsf{min}$ of the distances between $C$'s
constituent clusters and $C'$: Line 10 in
Algorithm~\ref{alg:cluster_cluster} now has $\Pone, \Ptwo$ jointly run
GC-protocol
$(\perp;Y)\leftarrow\minDist(R_{i,k}, R_{j,k}, X; B_{i,k}, B_{j,k})$
(see Appendix~\ref{app:sub})
to split the new distance
$\Delta_{i,k}=\min\{B_{i,k}-R_{i,k}, B_{j,k}-R_{j,k}\}$ into $X$,
$Y=\Delta_{i,k}+X$, without asymptotic efficiency changes.

More generally, the skeleton of protocol $\hcluster$ allows for
extensions that support a wider class of linkage functions, such as
average or centroid linkage, by appropriately refining GC-protocols
$\argMinSelect$, $\minDist$---but still, at quadratic cost per merged
cluster and cubic total cost. Yet, our single-linkage protocol can be
optimized to process each new cluster in only $O(\kappa n)$ time, for
a reduced $O(\kappa n^2)$ total running time, with GC-protocol
$(j;j)\leftarrow\argMinSelect (X;Y)$ now refined, on input arrays
$X,Y$, to return as common output the minimum-value index $j$ of
$Y-X$, excluding any non-linkage values.

The main idea is to exploit the associativity of operator
$\mathsf{min}$ and that single-linkage clustering only relates to minimum inter-cluster distances, to find the closest pair
$(i,j)$ in linear time, by looking up an array
$\bar{\Delta}=\bar{B}-\bar{R}$ storing the minimum row-wise
distances in $\Delta=B-R$ (a known technique in information
  retrieval~\cite[Section 17.2.1]{DBLP:books/daglib/0021593}).
Our modified protocol takes only $O(\kappa n)$
comparisons per clustering, as opposed to $O(\kappa n^2)$
of our main protocol. As shown in Section~\ref{sec:eval},
this results in significant performance improvement.


Specifically, at the end of the setup phase, $\Pone$, $\Ptwo$ now also
jointly run $(j_i;j_i)\leftarrow\argMinSelect (R_i; B_i)$,
$i\in[1:n]$, to learn the minimum-linkage index $j_i$ of the $i$th row
$B_i-R_i$ of $\Delta$ (excluding its $i$th location, as $B_{i,i}$,
$R_{i,i}$ store cluster $i$), and they both initialize array $\bar{J}$
as $\bar{J}_i= j_i$, whereas $\Pone$ initializes array $\bar{R}$ as
$\bar{R}_i=R_{i,j_i}$ and $\Ptwo$ array $\bar{B}$ as
$\bar{B}_i=B_{i,j_i}$.
Then, at the start of each iteration in the clustering phase (line 5
in Algorithm~\ref{alg:cluster_cluster}) and assuming that
$\bot = +\infty$, $\Pone$, $\Ptwo$ now jointly run
$(i;i)\leftarrow\argMinSelect (\bar{R}; \bar{B})$ to find the
closest-cluster pair $(i,j)$, $j=\bar{J}_i$, in only $O(\kappa n)$
time. Conveniently, as soon as they update linkages
$\Delta_{i,k}=\Delta_{k,i}$, for some $k\neq i, j$ (lines 9-12, as
$Y-X$ with
$(\perp;Y)\leftarrow\minDist(R_{i,k}, R_{j,k}, X; B_{i,k}, B_{j,k})$),
$\Pone$, $\Ptwo$ also update the joint state
$\{\bar{B}-\bar{R}, \bar{J}\}$ for updated row $m \in\{ i, k\}$:
First, by jointly running
$(z;z)\leftarrow\argMinSelect (\hat{R}; \hat{B})$ for arrays
$\hat{R}=[X, \bar{R}_m]$, $\hat{B}=[Y, \bar{B}_m]$ of size 2, and
then, if $z=1$, by setting $\bar{R}_m=\hat{R}_z$,
$\bar{B}_m=\hat{B}_z$ and $\bar{J}_m=\{i,k\}\setminus m$. At the end
of each iteration (lines 14-16), they also set
$\bar{R}_j=\bar{B}_j=\bar{J}_j=\bot$, as needed for consistency.

\smallskip\noindent {\bf Protocol extensions.} 
Our protocol can be easily adapted to handle higher dimensions ($d>1$). 
Its sub-protocol ($\hcluster.\cluster$) compares squared Euclidean distances 
thus it is almost unaffected by the number of dimensions; only the
setup and output phases need to be
modified, as follows. $\Ptwo$ computes helper
information $H$, representing each point not by 3 but by $3d$
encryptions (i.e., line 4 of Algorithm~\ref{alg:cluster_setup}
runs independently for each dimension). Analogously, $\Pone$, $\Ptwo$
compute square Euclidean distances (lines 3 and 11-12) as the sum
of squared per-dimension differences across all dimensions (over AHE). Shuffling 
remains largely unaffected, besides lists $L,S,S'$
consisting of $dn$ encryptions each. Finally,
representatives (line 6 in Algorithm~\ref{alg:cluster_filter}) are now
computed over vectors of $d$ values. Our protocol can also extended to other distance metrics,
e.g., $L_1$, $L_2$ or Euclidean, and any $L_p$ distance
for $p \ge 1$, with modifications for computing the 
distance matrix during setup. With squared Euclidean
the  distances are securely computed with
 AHE; for
other metrics, more
elaborate sub-protocol may be required.

\noindent {\bf Security.}  \iffull In
Appendix~\ref{app:proofs}, we prove the following result: \fi

\vspace{-0.15cm}
\begin{theorem}\label{THM:MAIN}
  Assuming Paillier's encryption scheme is semantically secure and
  that $\argMinSelect$ and $\maxDist$ are securely realized by
  GC-based protocols, protocol $\hcluster$ securely realizes
  functionality~$f^*_{HC}$.

\end{theorem}


\section{Scalability via approximation}\label{sec:approx}
The cryptographic machinery of our protocol imposes a noticeable overhead in practice. Although it is asymptotically similar to plaintext HC, standard operations are now replaced by cryptographic ones---no matter how well-optimized the code, such crypto-hardened operations will ultimately be slower.
Hence, to scale to larger datasets, we 
seek to exploit \emph{approximate} schemes for hierarchical clustering. In our case, approximation  refers to performing clustering over a high-volume  dataset by applying the $\mathsf{HCAlg}$ algorithm only on a small subset of the dataset. The effect of this is twofold: Cluster analysis is much faster but using fewer points lowers accuracy and increases  sensitivity to outliers.

In what follows, we adapt the $\CURE$ approximate clustering algorithm~\cite{cure} and seamlessly integrate it to our main protocol $\hcluster$, within a flexible design framework that offers a variety of configurations for balancing tradeoffs between performance and accuracy, to overall get the first variants of $\CURE$ for private collaborative hierarchical clustering. Although, in principle, our framework can be applied to any approximate clustering scheme (e.g., BIRCH~\cite{birch}), we choose $\CURE$ for its strong resilience to outliers and high accuracy (even on samples less than 1\% of original data)---features that place it among the best options for scalable hierarchical clustering.


\begin{figure}[t!]
\begin{minipage}[b]{6.5in}
	\fbox{
          \small
		\begin{minipage}{.5\textwidth}
                  \noindent\textbf{The $\CURE$ approximate clustering algorithm}

                  \smallskip\textbf{Input:} $\dataset , n,s,p,q,t_1,t_2,R$ \hspace*{\fill}\textbf{Output:} Clusters $\mathcal{C}$ over $\dataset$
                  
                  \smallskip\textbf{[Sampling]} Randomly pick $s$ points in $\dataset$ to form sample~$\mathcal{S}$.

                 \smallskip\textbf{[Clustering A]}

                 \smallskip~1. Partition $\mathcal{S}$ into $p$ partitions $\mathcal{P}_i$s, each of size $s/p$.

                 \smallskip~2. Run $\mathsf{HCAlg}$ to cluster each $\mathcal{P}_i$ into $s/(pq)$ target clusters.

                  \smallskip~3. Eliminate within each $\mathcal{P}_i$ clusters of size less than $t_1$. 

                  \smallskip\textbf{[Clustering B]}

                  \smallskip~1. Run $\mathsf{HCAlg}$ to cluster all remaining A-clusters $\mathcal{C}_A$ in $\mathcal{S}$.

                  \smallskip~2. Eliminate clusters of size less than $t_2$ to get B-clusters~$\mathcal{C}_B$.

                  \smallskip~3. Set $R$ random points in each B-cluster as its representatives.

                   \smallskip\textbf{[Classification]} 
                  
                  \smallskip~1.  Assign singletons in $\mathcal{D}$ to B-cluster of closest representative.
	
		\end{minipage}
	}
      \end{minipage}
\vspace{-0.4cm}

\caption{The $\CURE$ approximate clustering algorithm.}
\label{f:CURE}
\end{figure}

Described in Figure~\ref{f:CURE}, on input the original dataset $\dataset$ of size $n$ and a number of approximation parameters, $\CURE$ first randomly samples $s$ data points from $\dataset$ to form sample set $\mathcal{S}$. During A-clustering, $\mathcal{S}$ is partitioned into $p$ equally-sized parts $\mathcal{P}_1, \mathcal{P}_2, \ldots \mathcal{P}_p,$ and the ordinary algorithm $\mathsf{HCAlg}$ runs $p$ times to form a set $\mathcal{C}_A$ of A-clusters: Its $i$th execution is on input $\mathcal{P}_i$, $i\in [1,p]$, until exactly $s/(pq)$ clusters are formed, of which only those of size at least $t_1$ are included in $\mathcal{C}_A$ and the rest are eliminated as outliers. During B-clustering, $\mathsf{HCAlg}$ runs once again, this time over set $\mathcal{C}_A$, to form a set $\mathcal{C}_B$ of B-clusters, from which clusters of size less than $t_2>t_1$ are eventually eliminated as outliers. Finally, for each B-cluster in $\mathcal{C}_B$ a number of $R$ random representatives are selected, and each singleton point in $\mathcal{D}$ is included to the B-cluster containing its closest representative. Table~\ref{tab:cureparam} summarizes suggested values for each parameter as per $\CURE$'s original description~\cite{cure}.

\smallskip\noindent {\bf Private $\CURE$-approximate clustering.} We adapt the $\CURE$ algorithm to design private protocols for approximate clustering in our model for two-party joint hierarchical clustering. In applying our security formulation (Section~\ref{sec:def}) and our private protocols (Sections~\ref{sec:main}, ~\ref{sec:analysis}) to this problem instance, the following facts are vital:

\begin{itemize}[leftmargin=10pt]
\item[{\bf 1.}] $\CURE$ involves three main tasks: input \emph{sampling}, \emph{clustering} of sample, and unlabeled-points \emph{classification}.

\item[{\bf 2.}] Clustering involves $p+1$ invocations of $\mathsf{HCAlg'}$, which extends ordinary algorithm $\mathsf{HCAlg}$ to receive \emph{clusters as input} and compute its \emph{output over an input subset}. 

\item[{\bf 3.}]  If $p=1$ and $\mathcal{O}_A$, $\mathcal{O}_{B}$ are the A- and B-outliers, then:


  \noindent {\bf i.} $\mathsf{HCAlg'}$ first runs on $\mathcal{S}$ to form $\mathcal{C}_A$ over $\mathcal{S}_A\triangleq\mathcal{S}\setminus \mathcal{O}_A$; $\mathcal{C}_A$ is exactly the output of $\mathsf{HCAlg}$ run on $\mathcal{S}_A$; and next


  \noindent {\bf ii.} $\mathsf{HCAlg'}$ runs on $\mathcal{C}_A$ to form $\mathcal{C}_B$ over $\mathcal{S}_{B}\triangleq \mathcal{S}\setminus \{\mathcal{O}_A\cup \mathcal{O}_{B}\}$; $\mathcal{C}_B$ is exactly the output of $\mathsf{HCAlg}$ run on $\mathcal{S}_{B}$.
\end{itemize}

Fact 1 refines our protocol-design space to only securely realizing the clustering task, where sampling and classification are viewed as input pre-processing and output post-processing of clustering. Specifically, $\Pone$, $\Ptwo$: (1) individually form random input samples $\mathcal{S}_P$, $\mathcal{S}_Q$ of their own datasets $P, Q$; (2) compute B-clusters and their representatives (as specified by $\CURE$); and (3) use these B-cluster representatives to individually classify their own unlabeled points.

As such, the default private realization of $\CURE$ would entail having the parties perform \emph{clustering A and B jointly}. Yet, since our design space is already restricted to provide approximate solutions, we also consider two protocol variants, where parties trade even more accuracy for efficiency, by performing: (1) \emph{clustering A locally and only B jointly}; and, in the extreme case (2) \emph{clustering A and B locally}. We denote these protocols by $\jointFull$, $\jointSemi$ and~$\disjoint$.

In $\disjoint$, $\Pone$, $\Ptwo$ non-collaboratively compute B-clusters of their samples and announce the representatives selected. Though a degenerate solution, as it involves no interaction, this consideration is still useful: First, to serve as a baseline for evaluating the other variants, but mostly to further refine our design space. $\disjoint$ (trivially) preserves privacy during B-cluster computation, but violates the privacy guarantees offered by our point-agnostic dendrograms, by revealing a subset of a party's input points to the other party. To rectify this, present also in $\jointFull$ and $\jointSemi$, we fix $R=1$ and have each B-cluster be represented by its centroid. Using average values is expected to have no impact on accuracy, at least for spherical clusters (in~\cite{cure}, $R > 1$ is only used to improve accuracy of non-spherical clusters).
Fact 2 then ensures that B-clusters (and their centroids) can be computed by essentially running  algorithm $\mathsf{HCAlg}$, possibly with slight modifications (discussed below).

\begin{table}[t]
	\footnotesize
	\begin{center}
		\centering
		\begin{tabular}{c|c|c}
			Parameters & Description & Value \\
			\hline
			$n$, $s$ & Sizes of dataset and its sample & $\leq 1$M,  $[10^2 :10^3]$\\
			$p$, $q$ & \# parts, cluster/part control & $p=1,3,5$, $q=3$ \\
			$t_1$, $t_2$ & A-, B-cluster outlier thresholds & $ 3 = t_1 < t_2 = 5$ \\
			$R$ & Representatives per B-cluster & $R=1,3,5,7,10$
		\end{tabular}
	\end{center}
	\caption{$\CURE$ clustering parameters and values.}
	\label{tab:cureparam}
	\vspace{-0.4cm}
\end{table}

In $\jointSemi$, $\Pone$, $\Ptwo$ non-collaboratively compute A-clusters of their samples and then jointly merge them to B-clusters. Semantically,  they  run $\mathsf{HCAlg}$, not starting at level $n$ (singletons) but at an intermediate level $i$, where each input A-cluster contains at least $t_1$ points. Our $\hcluster$ can be employed, with one modification: At setup, the parties' joint state encodes their individual A-clusters and their pairwise linkages. Accordingly, sub-protocol $\hcluster.\setup$ is modified: (1) Lines 3 and 11 now compute inter-cluster distances (of same-party pairs),
and (2) lines 2 and 12 are used as a subroutine to compute all point distances across a given A-cluster pair, over which inter-cluster linkages (of cross-party pairs) are evaluated with $\argMinSelect$. The running time of modified $\hcluster.\setup$ is $O(\lambda s^2)$, as $O(s^2)$ distances are computed across $O(s)$ A-cluster points.

In $\jointFull$, $\Pone$, $\Ptwo$ jointly compute A- and B-clusters. This introduces the  challenge of how to transition from  A to B. Simply running $p$ copies of $\mathsf{HCAlg}$ in parallel for A-clusters does not provide the cluster linkages that are necessary for $\mathsf{HCAlg}$ to compute B-clusters. Possible solutions are either to treat A-clusters as singletons, which can drastically impair accuracy, or running an intermediate MPC protocol to bootstrap $\mathsf{HCAlg}$ with cluster linkages, which can impair performance. Instead, we simply fix $p=1$,  seamlessly using the final joint state of clustering A as initial  state for clustering B. Missing speedups by parallelism is  compensated by avoiding a costly bootstrap-protocol, at no accuracy loss, as our  experiments confirm ($p > 1$, is only suggested for parallelism in~\cite{cure}).

Finally, the security of protocols $\jointSemi$ and $\jointFull$ can be reduced to that of $\hcluster$. Our modular design and facts 2 and 3, ensure that security in our private $\CURE$-approximate clustering is captured by our ideal functionality $f^*_{HC}$ of Section~\ref{sec:def}: The intended two-party computation merely involves computing B-cluster representatives, which $f^*_{HC}$ provides, and any input/output modification in $\mathsf{HCAlg}$ causes a trivial change to the pre-/post-processing component of $f^*_{HC}$, consistent to our point-agnostic dendrograms. 


\section{Experimental evaluation}\label{sec:eval}



\textcolor{black}{Our main  goal is to evaluate the computational cost 
of our protocols and to determine the improvement of the optimized and approximate variants.} 
\textcolor{black}{We use four datasets from the UCI ML Repository~\cite{UCI}, restricted to numeric attributes: (1) $\iris$ for iris plants classification (150 records, 4 attributes); (2) $\wine$ for chemical analysis of wines (178 records, 13 attributes); (3) $\heart$ for heart disease diagnosis ($303$ records, $20$ attributes); and (4) $\bcancer$  for breast cancer diagnostics ($569$ records, $30$ attributes). As these are relatively small, we also generate our own synthetic datasets, scaling the size to millions of samples. Note that our protocol's performance depends mainly on the dataset size, is invariant to actual data values, and varies very little with data dimensionality, as our experiments confirm.}


\textcolor{black}{We introduced several variants of approximate clustering  based on  \CURE\  and want to evaluate their accuracy  and determine possible  between performance-accuracy tradeoffs. Traditionally, hierarchical clustering is an unsupervised learning task, for which accuracy metrics are not well defined. However, it is common to evaluate the accuracy of clustering via ground truth datasets including class labels on  samples. A good clustering algorithm will generate ``pure clusters'' and separate data according to the ground truth.  Each cluster will be labeled with the majority class of its samples, and the accuracy of the protocol is defined as the fraction of input points
that are clustered into their correct class relative to the ground truth. We employ this measure of accuracy to evaluate approximate clustering variants (\disjoint, \jointSemi, and \jointFull). Our standard privacy-preserving clustering protocol \hcluster\ and the optimized version \opt~maintain the same accuracy as the original non-private protocol, hence we do not report accuracy for them.}


{
\centering
\begin{figure}[t!]
	\begin{minipage}[b]{.5\textwidth}
	\begin{subfigure}[t]{0.51\columnwidth}
         \includegraphics[width=\textwidth]{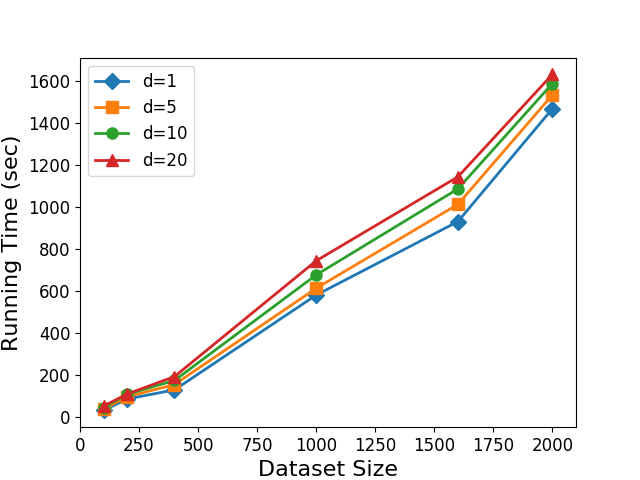}
          \vspace{-0.7cm}
          \caption{\scriptsize Computation cost of \base.}\label{fig:crypto-time}
	\end{subfigure}
	\begin{subfigure}[t]{0.51\columnwidth}
          \includegraphics[width=\textwidth]{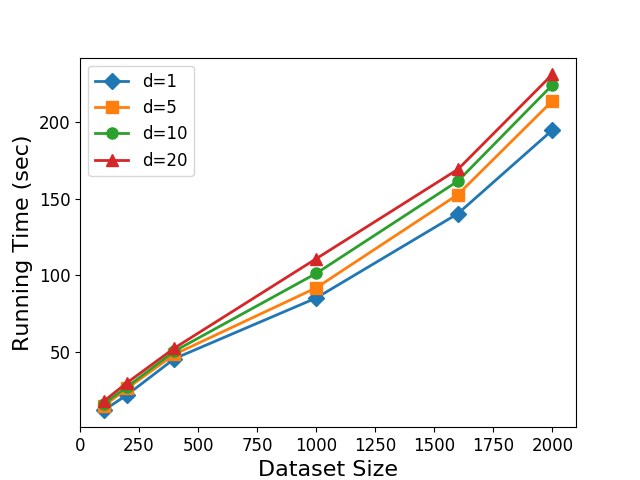}
          \vspace{-0.7cm}
          \caption{\scriptsize Computation cost of \opt.}\label{fig:crypto-time-op}
	\end{subfigure}
      \end{minipage}
     \begin{minipage}[b]{.5\textwidth}
	\begin{subfigure}[t]{0.49\columnwidth}
         \includegraphics[width=\textwidth]{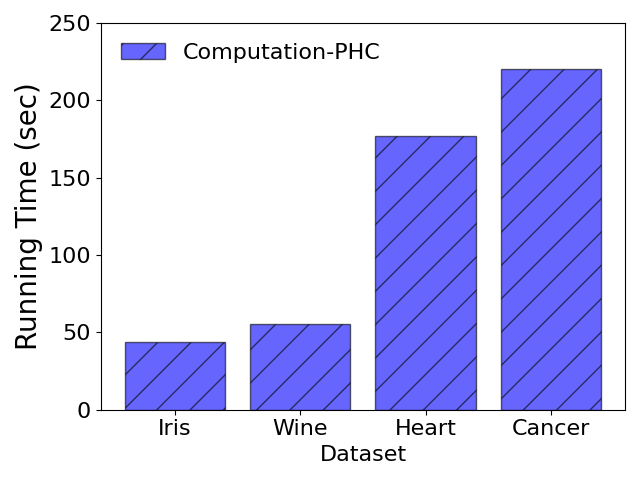}
          \vspace{-0.7cm}
          \caption{\scriptsize Real-data performance of \base.}\label{fig:crypto-real}
	\end{subfigure}
	\begin{subfigure}[t]{0.49\columnwidth}
          \includegraphics[width=\textwidth]{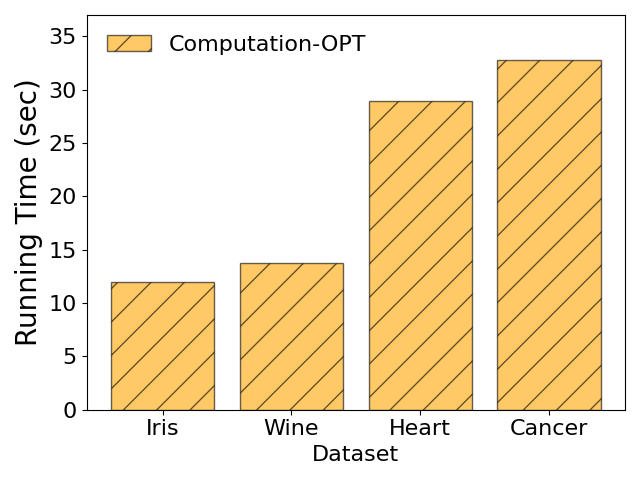}
          \vspace{-0.7cm}
          \caption{\scriptsize Real-data performance of \opt.}\label{fig:crypto-real-op}
	\end{subfigure}
      \end{minipage}
  	\vspace{-0.5cm}
      \caption{Performance of \base~(left) Vs. \opt~(right).}\label{fig:crypto}
\end{figure}
}

\Comment{
Figure~\ref{fig:exdata} provides a visualization of one sample dataset generated by our method above.

\begin{figure}[h!tp]
	\centering
    \vspace{-0.4cm}
	\includegraphics[width=.6\columnwidth]{graphs/fig_data.png}
		\vspace{-0.4cm}
	\caption{Synthetic data of 100K records and 0.1\% outliers. }
	\label{fig:exdata}
      \end{figure}
      }

We generate synthetic $d$-dimensional datasets of sizes up to 1M records and $d \in [1,20]$, using a Gaussian mixture distribution, as follows: (1) The number of clusters is randomly chosen in $[8:15]$; (2) Each cluster center is randomly chosen in $[-50,50]^d$ (performance is dominated by $\kappa$ but not exact data values), subject to a minimum-separation distance between pairs; (3) Cluster standard deviation is randomly chosen in $[0.5,4$]; and (4) Outliers are selected uniformly at random in the same interval and assigned randomly to clusters to emulate 3 noise percentage scenarios: low $0.1\%$, medium $1\%$, and high~$5\%$.
We randomly split each dataset into two halves which form the private inputs of the parties. We set the number of target clusters to $\ell_t=5$; as our protocol incurs costs linear in the number of iterations ($n-\ell_t$), this choice comprises a worst-case setting, as in practice more than 5 target clusters are desired.

We adapted our protocols to support floating point numbers. 
Here, due to the simplicity of the involved operations,
we can rely on fixed-precision floating point numbers and  it suffices to multiply floating point values by a constant $K$ (e.g. $K = 2^{20}$ for IEEE 754 doubles). 
During $\hcluster.\setup$, we can achieve higher precision. After each party decrypts the blinded values (line~\ref{line:decrypt1} and line~\ref{line:decrypt2}), they can re-scale by dividing the constant $K$ without affecting precision.
During $\cluster$, as we only merge the points 
based on the comparisons between the distances,  multiplying by a constant does not affect the results.

Finally, we use the ABY C++ framework~\cite{ABY}, $128$-bit AES for GC, $1024$-bits Pailler, and set $\kappa =40$.  We use \texttt{libpaillier}~\cite{libpaillier} for Paillier encryption.  We run our experiments on a 24-core machine, running Scientific Linux with 128GB memory on 2.9GHz Intel Xeon.

\begin{figure}[t!]	
	\begin{subfigure}[t]{0.48\columnwidth}
		\includegraphics[width=\textwidth]{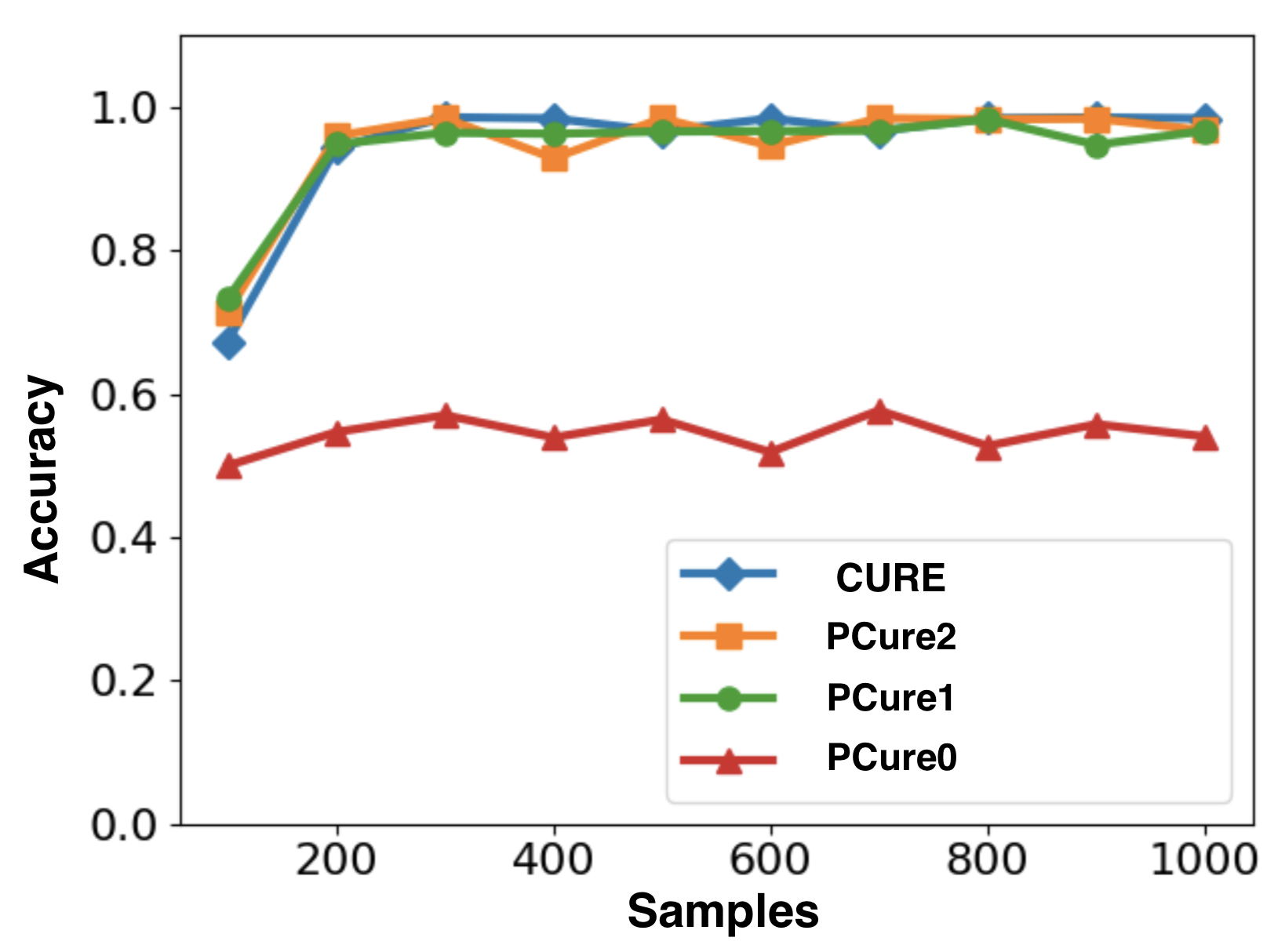}
	\end{subfigure}
	\begin{subfigure}[t]{0.48\columnwidth}
		\includegraphics[width=\textwidth]{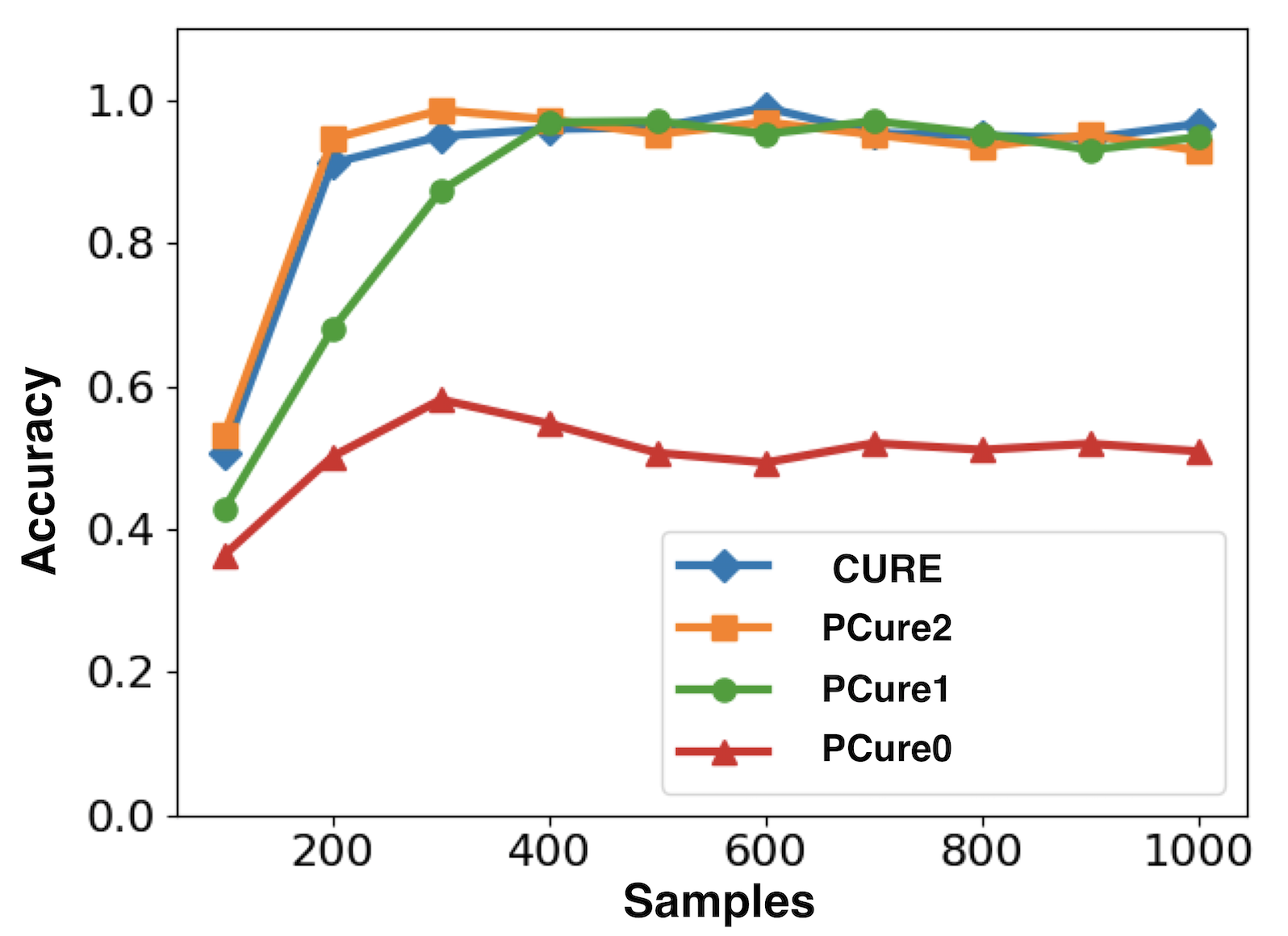}
	\end{subfigure}
	\begin{subfigure}[t]{0.48\columnwidth}			\includegraphics[width=\textwidth]{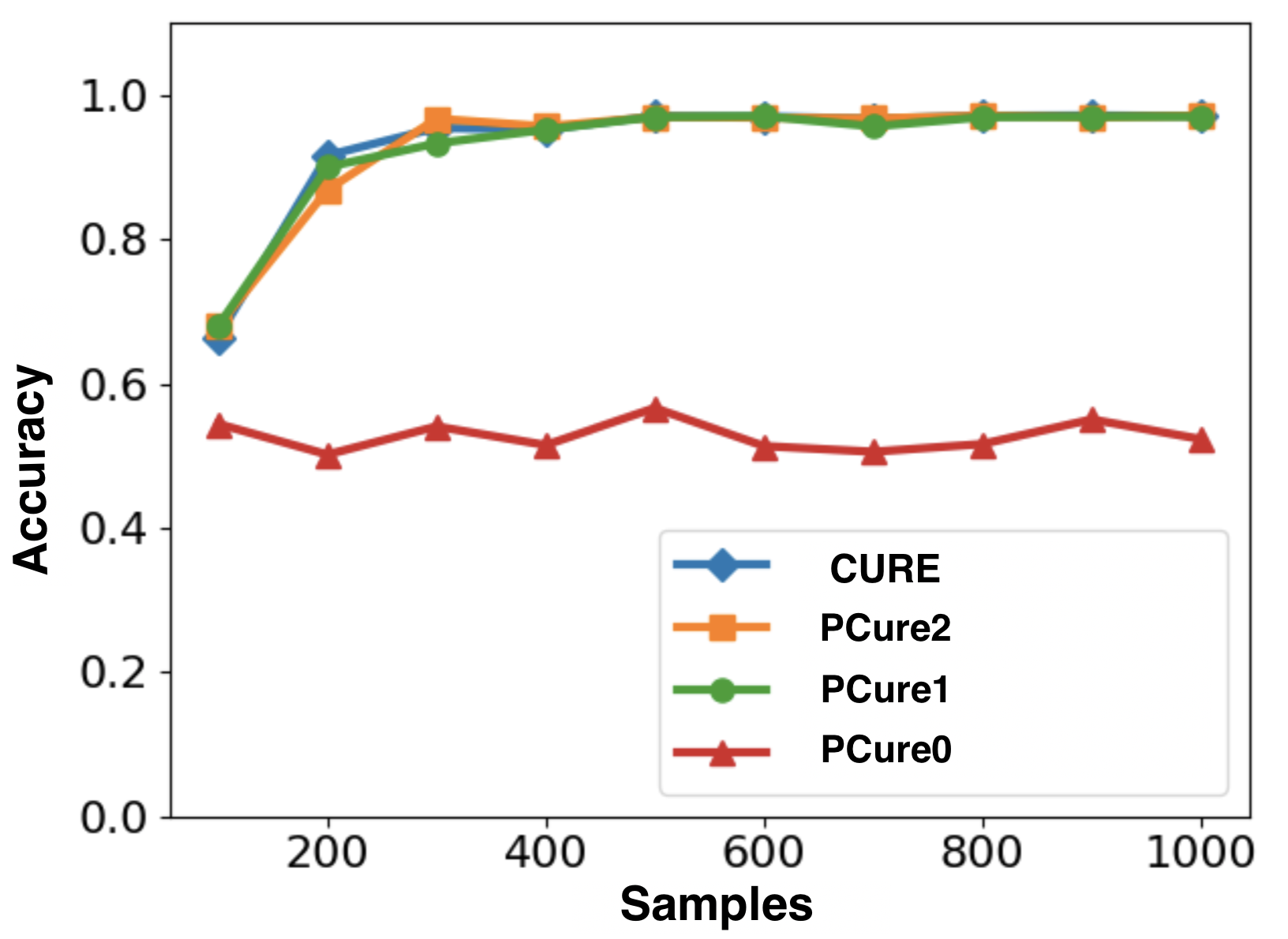}
	\end{subfigure}
	\begin{subfigure}[t]{0.48\columnwidth}
		\includegraphics[width=\textwidth]{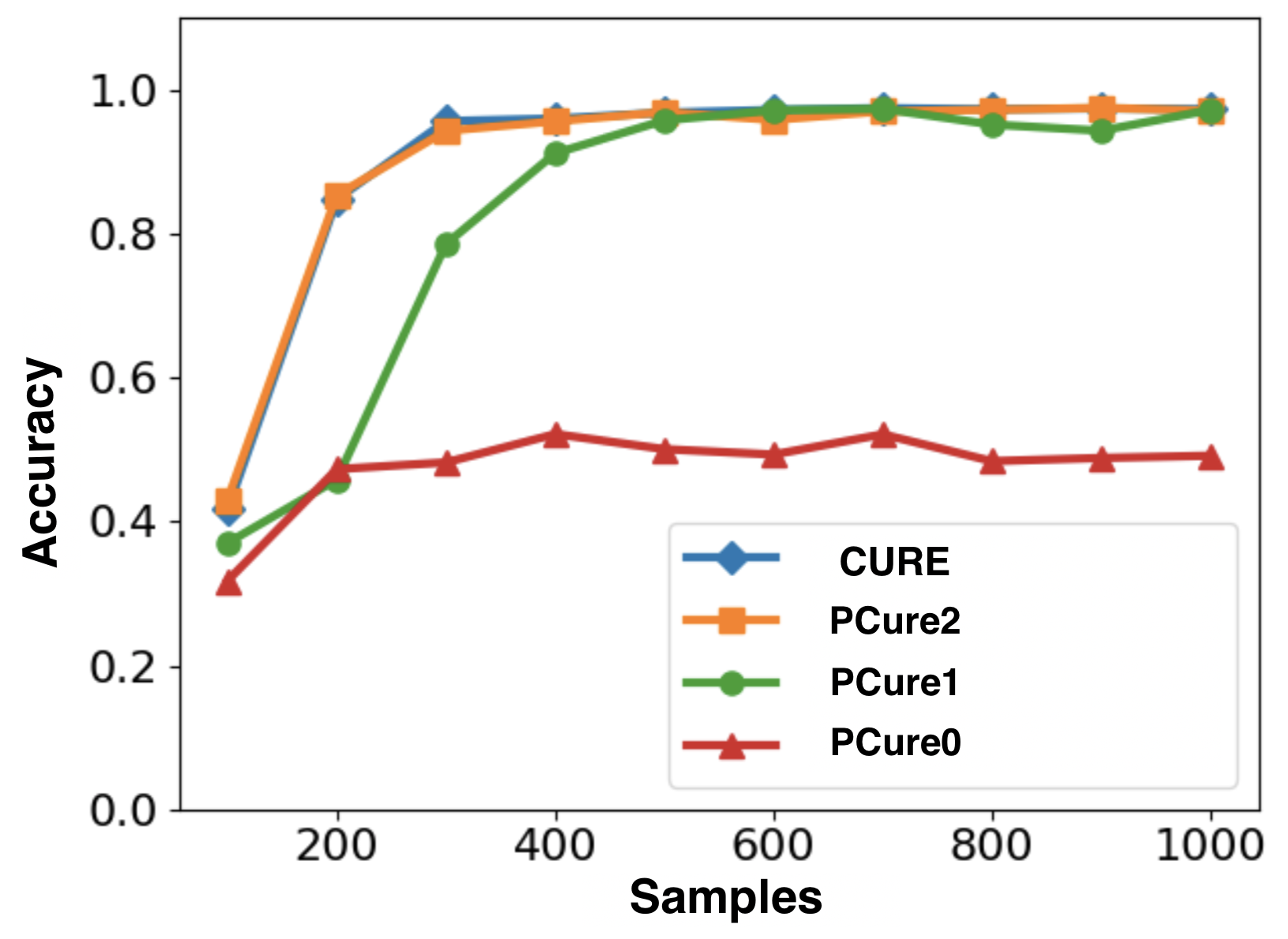}
	\end{subfigure}
	\begin{subfigure}[t]{0.48\columnwidth}				\includegraphics[width=\textwidth]{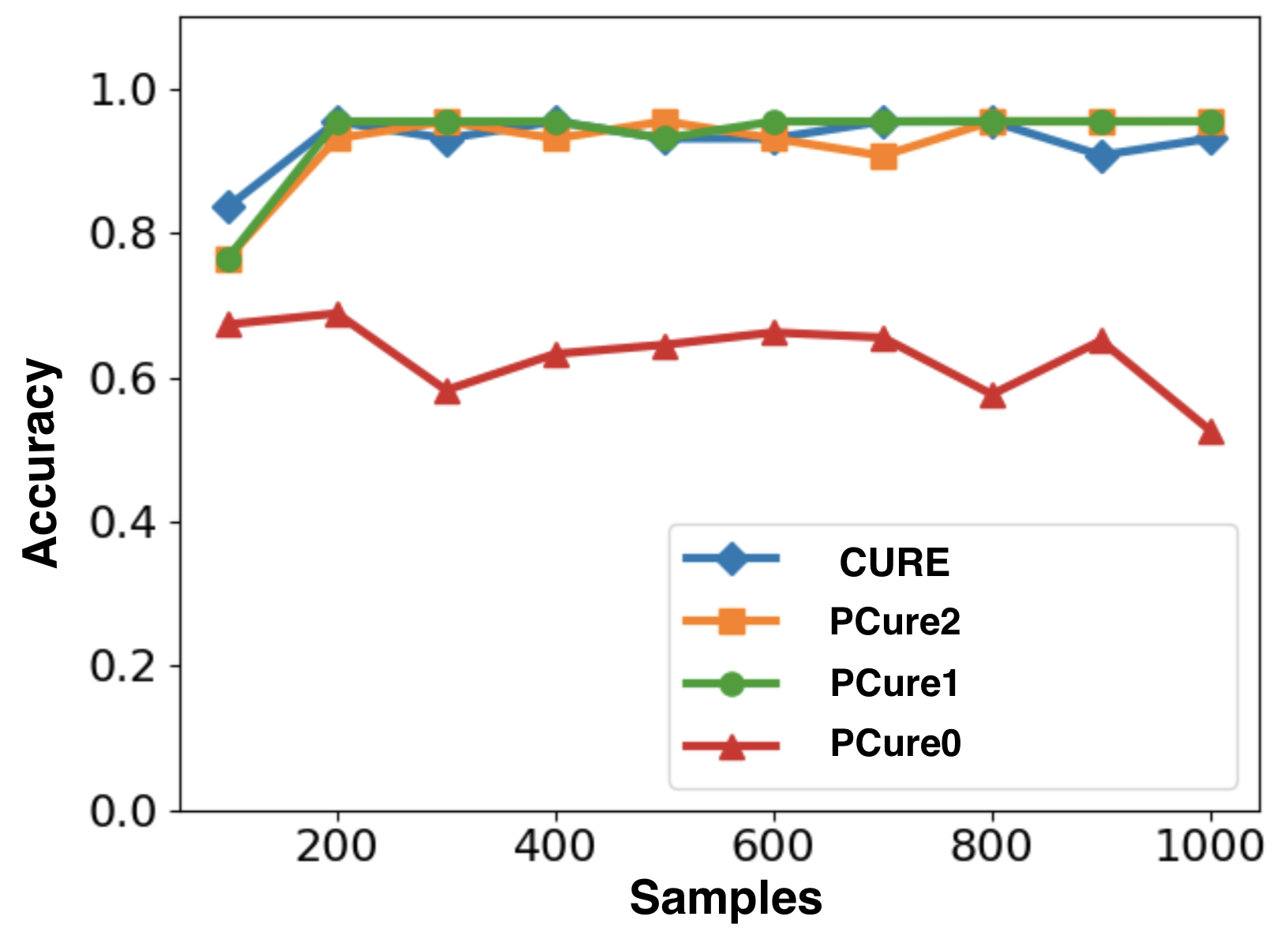}
	\end{subfigure}
	~
	\begin{subfigure}[t]{0.48\columnwidth}
		\includegraphics[width=\textwidth]{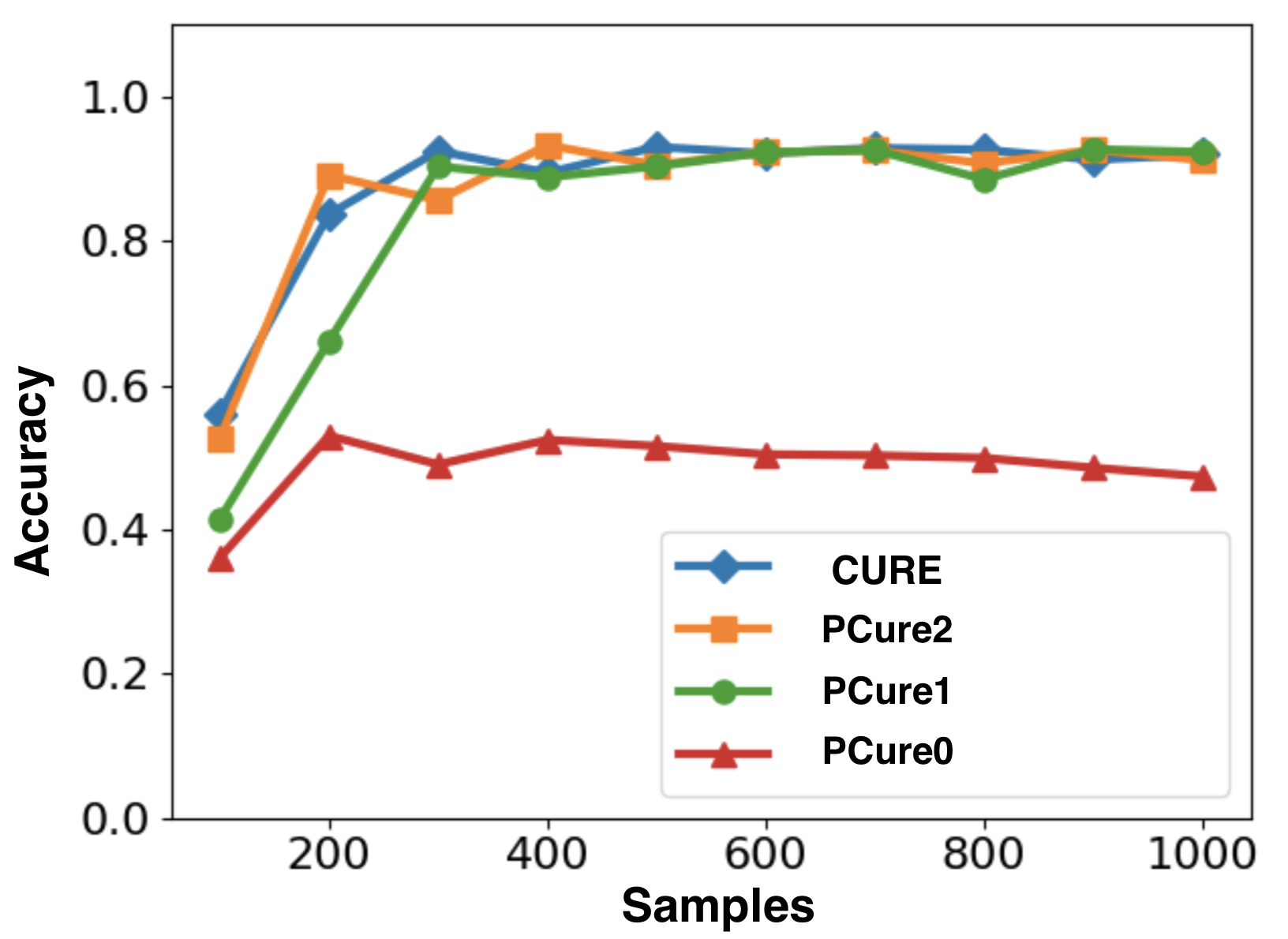}
	\end{subfigure}
		\vspace{-0.4cm}
	\caption{Accuracy of $\CURE$, $\mathsf{PCure*}$: $p=1$ (left), $p=5$ (right), \#outliers = $0.1\%$~(top), $1\%$~(middle), $5\%$~(bottom).}
	\label{fig:1millionapprox}
\end{figure}

\noindent {\bf Protocol $\hcluster$.} We first report results on the performance of our $\hcluster$ protocol from Section~\ref{sec:main}. Figure~\ref{fig:crypto-time}
shows the computational cost for synthetic datasets of various sizes and dimensions, averaged over single and complete linkages. First, consistently with our analysis in Section~\ref{sec:analysis}, dimensions have minimal impact, since $\hcluster$'s performance relates primarily to computing inter-cluster distances that is minimally affected by $d$. As expected its cubic asymptotic complexity, the overhead increases steeply with dataset size $n$.

\smallskip
 
\noindent {\bf Protocol $\opt$.} 
Figure~\ref{fig:crypto-time-op} 
shows the computational costs on synthetic datasets for our optimized single-linkage variant $\opt$
(with  configurations identical to those for $\hcluster$).  In line with our analysis in Section~\ref{sec:analysis}, $\opt$ significantly improves performance,  reducing running time by an order of magnitude. E.g., for datasets of $2000$ $20$-dimensional points, the running time is approximately $230$ secs, an $8\times$ speedup compared to \base. 
The difference 
in our above example,is explained by the following observations: (1) although $\opt$ improves performance during clustering by a linear factor, it adds costs during setup; and (2) the involved constants of the quadratic costs are higher for running time
in setup phase, and vice versa in clustering phase.
As  shown in Figure~\ref{fig:crypto-real-op}, 
$\opt$ significantly improves performance over \base, also when tested over our real datasets.

\smallskip\noindent {\bf Protocols $\mathsf{PCure*}$.} 
Figure~\ref{fig:1millionapprox} shows the accuracy of our $\CURE$-variant protocols $\disjoint$, $\jointSemi$, $\jointFull$ from Section~\ref{sec:approx}, and the non-private $\CURE$ algorithm on synthetic datasets of 1M records for $10^2$--$10^3$ samples, partition parameters $p=1$ and $p=5$, and for low (0.1\%), medium (1\%), and high (5\%) outliers-to-data percentages.
Clearly, $\disjoint$, where parties run $\CURE$ on their own samples, without any interaction besides announcing representatives for individually computed clusters, exhibits very poor accuracy.
e.g., $44.4$\% loss for 1M records.
For $p=1$, $\jointSemi$ and $\jointFull$ achieve similar accuracy, which approaches that of $\CURE$ for large enough samples: At 300 samples or higher, the gap is within 3\%. For higher values of $p$, e.g., $p=5$, $\jointSemi$ and $\jointFull$ exhibit a difference in accuracy: E.g., at 200 samples the accuracy for $\jointSemi$ is lower by 39.54\% than $\jointFull$; but at 500 samples or more, they are within~3.18\%.

Moreover, experimenting with all combinations of $p=1,3,5$ partitions and $R=1,3,5,7,10$ representatives shows that the accuracies of $\jointFull$ and $\jointSemi$ are very close to $\CURE$ at $s=1000$ samples (or more). The largest observed difference between $\jointSemi$ and $\CURE$ is 3.57\%, and between $\jointFull$ and $\CURE$ is 2.7\%.  For $p=1$ and $R=1$ either difference is less than 1\% at 1000 samples (or more). Thus, our choice of $p=1$ and $R=1$ to protect data privacy, as argued in Section~\ref{sec:approx}, does not impact the protocol's accuracy.

\begin{figure}[t!]
	\begin{subfigure}[t]{\columnwidth}
		\centering
		\includegraphics[width=.6\textwidth]{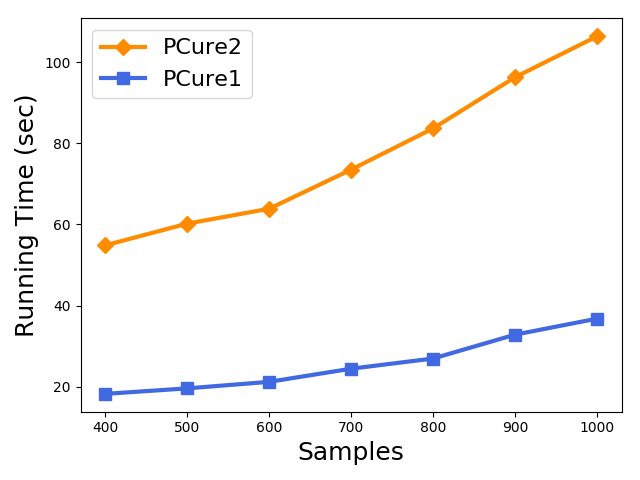}
	\end{subfigure}
	\vspace{-.9cm}
	\caption{End-to-end computation of $\jointSemi$, $\jointFull$.}
	\label{fig:e2e-comp}
\end{figure}

\textcolor{black}{
We also compare end-to-end computation for  $\jointSemi$ and $\jointFull$ (with \opt), $n=10^6$, $d=10$, $1\%$ outliers, no sample partitioning ($p=1$), $\ell_t=5$ target clusters, and $q=3$ for A-Clustering.
Figure~\ref{fig:e2e-comp} shows their good performance 
for sample sizes $s\in[400:1000]$. For $10^3$ samples, $\jointFull$ runs in $104$sec, while $\jointSemi$ runs in $35$sec -- $3\times$ faster, but with similar accuracy 97.09\%.}

\let\oldsim\sim 
\renewcommand{\sim}{{\oldsim}}

\smallskip\noindent \textcolor{black}{
{\bf Network Latency Impact.}
Although our experiments show the efficiency of our schemes, if executed over WAN this would be affected by network latency and data transmission. To estimate this impact, we considered two AWS machines in \textsf{us-east} and \textsf{us-west} and measured their latency to $50-60$ms. Taking $\jointFull$ with $400$ samples and $\ell_t=5$ as our use case, a single clustering round  with four roundtrips (assuming  distance update is done with a single garbled circuit) would take approximately $200$-$240$ms. Regarding data transmission of the two garbled circuits for finding the minimum distance and updating the cluster distances, using the circuits for addition/subtraction, comparison, and min-index-selection from~\cite{min1} for $100$, $64$-bit values, we estimate their size as roughly $10$MB (not including the OT data which is dominated by the circuits). Under the mild assumption of a $100$Mbps connection, transmission would take $\sim800$ms for a total of $\sim1$sec. In subsequent rounds, the circuits become progressively smaller but the number of roundtrips remains the same; even conservatively multiplying by $395$ rounds, we have approximately $400$sec of total communication time. For comparison, in Figure~\ref{fig:e2e-comp}, for the same setting computation takes $\sim55$sec.}

\textcolor{black}{
Hence, communication indeed becomes a bottleneck for our schemes when run over WAN, but not to the point where they are entirely impractical. Furthermore,  our goal when implementing our schemes was not to minimize end-to-end latency but computation, so there is plenty of room for optimizations. E.g., our protocols can be run in ``round batches'' merging $k$ clusters with one interaction (by larger circuits) which would decrease RTTs by a factor of $k$.  Finally, dedicated cloud technologies, such as AWS VPC~\cite{vpc}, can offer private connections
drastically reducing communication time.}

\section{Related Work}
\label{sec:related}


\smallskip\noindent {\bf Secure machine learning.}
There exists a rapidly growing line of works that propose secure protocols for a variety of ML tasks. This includes constructions for private classification models in the supervised learning setting (such as decision trees~\cite{PPDM}, SVM classification~\cite{PPSVM}, linear regression~\cite{PPLIN1,PPLIN2,PPLIN3}, logistic regression~\cite{PPLOG} and neural networks~\cite{nn1,ecg,nn2}), as well as federated learning tasks~\cite{Bonawitz17}. Another focus has been on proposing MPC-based protocols that are provably secure under a well-defined real/ideal definition, similar to ours (e.g.,~\cite{popa,PPRidge,PPRidge2,gilad,aono, define, opt1,opt2,opt3,opt4,opt5,CryptoNets,secureML,opt6,opt7}), for numerous tasks with a focus on neural networks and deep learning.

The above works can be split into two categories: those that focus on private  model training 
and those that focus on private inference/classification. 
In our unsupervised setting, our protocol protects the privacy of the parties' data during the clustering phase.



\smallskip\noindent {\bf Deployed techniques.}
In terms of techniques, most works use (some variant of) homomorphic encryption (e.g.,~\cite{Paillier99,FHE}). More advanced ML tasks often require hybrid techniques, e.g., combining the above with garbled circuits (e.g.,~\cite{opt2}) or other MPC techniques~\cite{opt1,delphi}. Our construction adopts such ``mixed'' techniques for the problem of hierarchical clustering. More recently, solutions have been proposed based on trusted hardware (such as Intel SGX), e.g.,~\cite{olga,slalom,ntee}. This avoids the need for ``heavy'' cryptography, however, it weakens the threat model as it requires trusting the hardware vendor. \textcolor{black}{Finally, a different approach 
seeks privacy via data perturbation~\cite{dp1,dp2,dp3,dp4,perturb,deep1}, by adding statistical noise to
hide data values, e.g., differential privacy~\cite{dmns}. Such
techniques are orthogonal to the cryptographic methods that we apply here but they can potentially be combined (e.g., as in~\cite{dpmpc1}). Using noise to hide whether a specific point has been included in a given cluster would be complement very nicely our cluster-information-reduction approach, potentially leading to more robust security treatment.}

\smallskip\noindent {\bf Privacy-preserving clustering.}
Many previous works proposed private solutions for different clustering tasks with the majority focusing on the popular, but conceptually simpler, $k$-means problem (e.g.,~\cite{kk_original,kk1,Jha05,kk2,kk4,kk5,nicek1,nicek2,nicek3,nicek4}) and other partitioning-based clustering methods (e.g.,~\cite{nikosother,nikosother2}). Fewer other works  consider private density-based~\cite{dense1,dense2,dense3} or distribution-based~\cite{distro1} clustering. An in-depth literature survey and comparative analysis of private clustering schemes can be found in the recent work of~\cite{hegde}. 

Focusing on private hierarchical clustering, no previous work offers a formal security definition, relying instead on ad-hoc analysis~\cite{pphc3,pphc5,pphc1,PHC14,another}. Moreover some schemes leak information to the participants that can clearly be harfmul---and is much more than what our protocol reveals---e.g.,~\cite{new1,new2} reveal {all distances across parties' records}. One notable exception is the scheme of Su et al.~\cite{goodpphc} which, however, is designed specifically for the case of document clustering. Here, we proposed a security formulation within the widely studied read/ideal paradigm of MPC that characterizes precisely what information is revealed to the collaborating parties. Besides making it  easier to compare our solution with potential future ones that follow our formulation, this is, to the best of our knowledge the only private hierarchical clustering scheme with formal proofs of security.
Finally, it is an interesting problem to combine optimizations for ``plaintext'' clustering (e.g.,\cite{plain1,plain2,plain3}) with privacy-preserving techniques to improve efficiency. 

\smallskip\noindent {\bf Secure approximate computation.}
The interplay between cryptography and efficient approximation~\cite{approx1} has already been studied for pattern matching in genomic data~\cite{approx2,approx3}, $k$-means~\cite{kk3}, and logistic regression~\cite{approx4,approx5}. To the best of our knowledge, ours is the first work to compose secure cryptographic protocols with efficient approximation algorithms for hierarchical clustering.

\smallskip\noindent {\bf Leakage in machine learning.}
The significant impact of information leakage in collaborative, distributed, or federated learning has been the topic of a long line of research (e.g., see~\cite{deep1,10.1145/3436755,flsurvey,surv}). Various practical attacks have been demonstrated that infer information about the training data or the ML model and its hyper-parameters, (e.g.,~\cite{ModelInversion15,GAN17,MembershipInversion17}).
This is even more important in collaborative learning where parties  could otherwise benefit from sharing data but such leakage may stop them (e.g.,~\cite{leak1,leak2,leak3,leak4}). Hence, it is crucial for our protocol to formally characterize what is the shared information for the two parties.


\section{Conclusion}
\label{sec:con}
We propose the first formal security definition for private hierarchical clustering and design a protocol for single and complete linkage, as well as an optimized version. We also combine this with approximate clustering to increase scalability.
\textcolor{black}{We hope this work motivates further research in privacy-preserving unsupervised learning, including secure protocols for other linkage types (e.g., Ward), alternative approximation frameworks (e.g., BIRCH~\cite{birch}), different tasks (e.g., mixture models, association rules or graph learning), or schemes for more than two parties to benefit from larger-scale collaborations. Specific to our definition of privacy, we believe it would be  helpful to experimentally and empirically evaluate the impact (even our significantly redacted) dendrogram leakage can have, e.g., by demonstrating possible leakage-abuse attacks.}


\Comment{
Extensions towards maintaining security also in the malicious threat
model, where parties may misbehave arbitrarily, are generally
possible---though, with new design challenges emerged. Techniques that
can be applied include augmenting homomorphic ciphertexts with
zero-knowledge proofs and garbled circuits with cut-and-choose
extensions, as well as, recent developments in efficient secure
computation~\cite{DBLP:conf/ccs/WangRK17}.

Extending our protocol to support multiple participants, as typically
considered by federated learning, is our main direction for future
work. We believe that this goal is feasible but would require
drastically different techniques to achieve practical
performance. Techniques that may help in this direction include
replacing the garbled circuit components with secret-sharing based
protocols~\cite{Araki16} and using recent advancements in secure
computation~\cite{global,overdrive}.
}






\begin{acks}
The authors would like to thank the members of the AWS Crypto team for their useful comments and inputs, the anonymous reviewers for their valuable feedback, and Anrin Chakraborti for shepherding this paper.  Dimitrios Papadopoulos was supported by the Hong Kong Research Grants Council (RGC) under grant ECS-26208318. Alina Oprea and Nikos Triandopoulos were supported by the National Science Foundation (NSF) under grants CNS-171763 and CNS-1718782.

\end{acks}

\balance
\bibliographystyle{ACM-Reference-Format}
\bibliography{alina,nikos}

\appendix
\section{Garbled circuits}\label{app:gc}

Garbled circuits (GC)~\cite{Yao1,Yao2} provide a general framework for
securely realizing two-party computation of any functionality. The
framework has been thoroughly studied in the literature (e.g., see
formal treatments of the topic~\cite{GC1,GC2}) and we here
overview the specific procedures involved in it.

In our running example, parties $\Pone$ and $\Ptwo$ wish to evaluate a
specific function $f$ over their respective inputs $x_1,x_2$ and
engage in an interactive 2-phase protocol, where one party plays the
role of the \emph{garbler} and the other the role of the
\emph{evaluator}. Without loss of generality, $\Pone$ is the garbler
and $\Ptwo$ is the evaluator, and their interaction proceeds as
follows.

In phase I, $\Pone$ expresses $f$ as a Boolean circuit $\mathcal{C}_f$,
i.e., as a directed acyclic graph of Boolean AND and OR gates, and
then sends a ``garbled,'' i.e., encrypted, version of $\mathcal{C}_f$ to
$\Ptwo$.

\begin{figure}[b]
	\hspace{0.3cm}\includegraphics[width=0.9\linewidth]{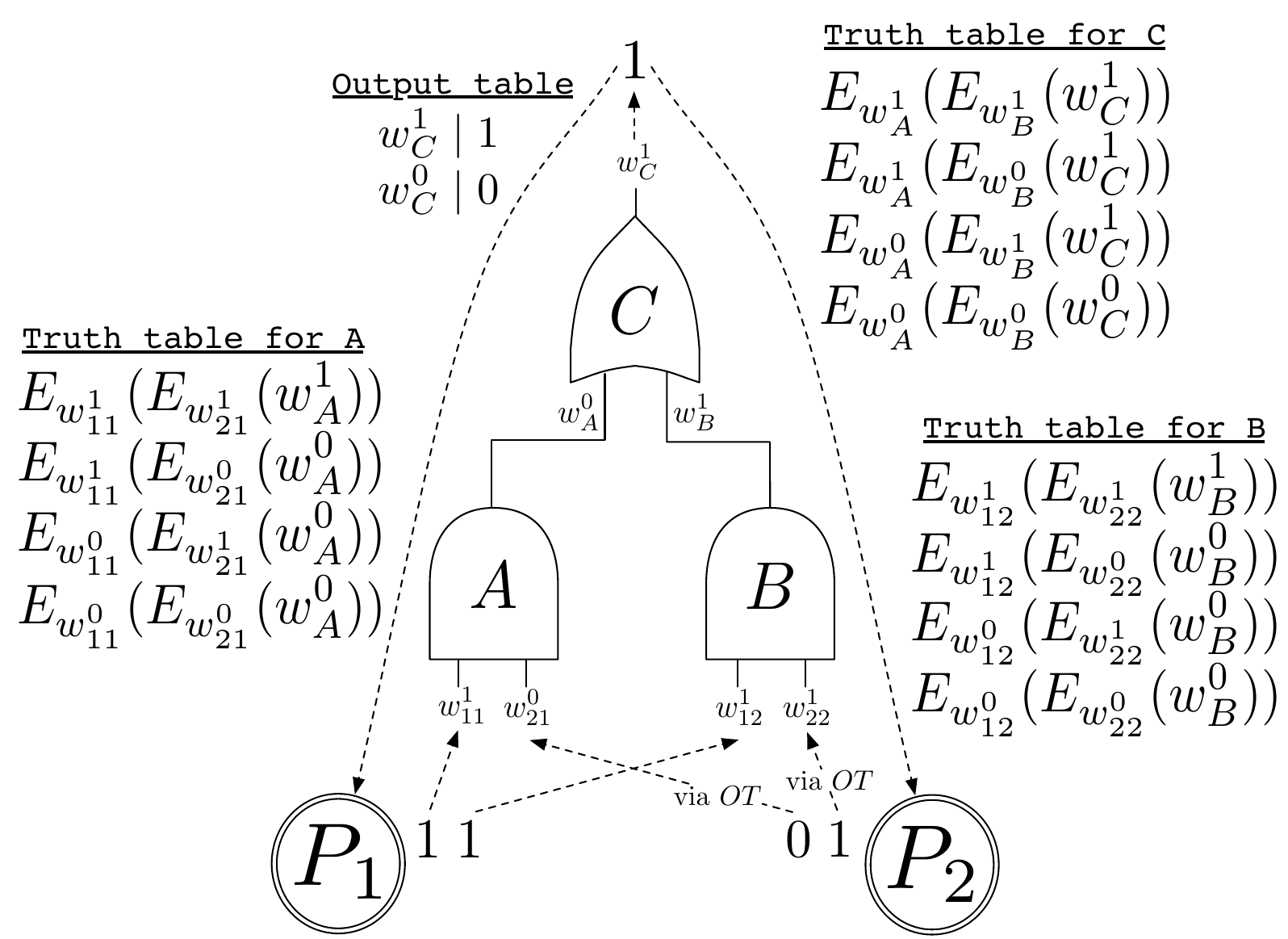}
	\caption{Garbled circuit $\mathcal{C}_f$ of a specific
		function $f$ that computes the OR over the pairwise ANDs of
		the 2-bit inputs.}\label{fig:garbled_circuit}
\end{figure}

In our example, $\mathcal{C}_f$ corresponds to a circuit of two AND
gates $A,B$ and an OR gate $C$, shown in
Figure~\ref{fig:garbled_circuit}: Inputs $x_1$, $x_2$ are $11$ and
$01$, and output $f(x_1,x_2)$ is 1, computed by feeding to the OR
gate the two bitwise ANDs of the inputs.

To garble $\mathcal{C}_f$, $\Pone$ first maps (the two possible bits
0, 1 of) each wire $X$ in $\mathcal{C}_f$ to two random values
$w^0_X, w^1_X$ (from a large domain, e.g., $\{0,1\}^{128}$), called
the \emph{garbled values} of~$X$.

Specifically, $\Pone$ maps the output wires of gates $A$, $B$, and $C$
to random garbled values $\{w_{A}^0, w_{A}^1\}, \{w_{B}^0, w_{B}^1\}$
and respectively $\{w_{C}^0, w_{C}^1\}$, and also maps the two input
wires of gate $A$ (respectively, gate $B$) to random garbled values
$\{w_{11}^0, w_{11}^1\}, \{w_{21}^0, w_{21}^1\}$ (respectively,
$\{w_{12}^0, w_{12}^1\}, \{w_{22}^0, w_{22}^1\}$), where mnemonically
the $i$-th input bit of party $\mathsf{P}_j$ corresponds to the
$ij$-wire, $i,j\in \{1,2\}$.

Next, $\Pone$ sends to $\Ptwo$ the \emph{garbled truth table} of every
Boolean gate in $\mathcal{C}_f$, which is the \emph{permuted}
encrypted truth table of the gate, where row in the truth table is
appropriately encrypted using the garbled values of its three
associated wires. We only specify the garbled truth table of the AND
gate $A$, as other gates can be handled similarly. The row
$(1,1) \rightarrow 1$ in the truth table of $A$ dictates that the
output is $1$ when input is $1,1$ or, using garbled values, that the
output is $w_{A}^1$ when input is $w_{11}^1, w_{21}^1$. Accordingly,
using a semantically-secure symmetric encryption scheme $E_{k}(\cdot)$
(e.g., 128-bit AES), $\Pone$ can express this condition as ciphertext
$E_{w_{11}^1}(E_{w_{21}^1}(w_{A}^1))$, where the output $w_{A}^1$ is
\emph{successively} encrypted using the inputs $w_{11}^1, w_{21}^1$ as
encryption keys. $\Pone$ produces a similar ciphetext for each other
row in the truth table of $A$ and sends them to $\Ptwo$,
\emph{permuted} to hide the order of the rows. Observe that one can
retrieve $w_{A}^1$ if and only if they possess \emph{both}
$w_{11}^1, w_{21}^1$, and that if one possesses \emph{only}
$w_{11}^1, w_{21}^1$, all other entries of the garbled truth table of
$A$ (besides $w_{A}^1$) are indistinguishable from random, due to the
semantic security of the encryption scheme $E_{k}(\cdot)$.

Finally, to allow $\Ptwo$ to retrieve the final output $f(x_1,x_2)$,
$\Pone$ also sends the garbled values $w_C^0,w_C^1$ of the output wire
\emph{together} with their corresponding mapping to 0 and 1. Note that
$\Ptwo$ is no privy to any other mappings between wires' garbled
values and their possible bit values.

In phase II, $\Ptwo$ evaluates the entire circuit $\mathcal{C}_f$ over
the received garbled truth tables of the gates in it, by evaluating
gates one by one in the ordering hierarchy induced by (the DAG
structure of) $\mathcal{C}_f$. Indeed, if $\Ptwo$ knows the $w$ value
of each input wire of a gate and its garbled truth table, $\Ptwo$ can
easily discover its output value, by attempting to decrypt all rows in
the table and accepting only the one that returns a correct output
value. For example, if $\Ptwo$ has $w_{11}^1, w_{21}^0$, $\Ptwo$ can
try to decrypt every value in the garbled truth table of $A$, until
$\Ptwo$ finds the correct value $w_{A}^1$.\footnote{\scriptsize For
  this, we need to assume that the encryption scheme allows detection
  of well-formed decryptions, i.e., it is possible to deduce whether
  the retrieved plaintext has a correct format. This can be easily
  achieved using a blockcipher and padding with a sufficient number of
  $0$'s, in which case well-formed decryptions will have a long suffix
  of $0$'s and decryptions under the wrong key will have a suffix of
  random bits. This property is referred to as \emph{verifiable range}
  in~\cite{GC1}.}

To initiate this circuit-evaluation process, $\Ptwo$ needs to learn
the garbled values of each of the input wires in $\mathcal{C}_f$,
which is achieved as follows: (1) $\Pone$ sends to $\Ptwo$ the $w$
values $w_{11}^1, w_{12}^1$ corresponding to the input wires of
$\Pone$ in the clear (note that since these are random values, $\Ptwo$
cannot map them to 0 or 1, thus $\Pone$'s input is protected); (2)
$\Ptwo$ privately query from $\Pone$ the $w$ values
$w_{21}^0, w_{22}^1$ corresponding to the input wires of $\Ptwo$, that
is, without $\Pone$ learning which garbled values were queried, via a
two-party secure computation protocol called 1-out-of-2
\emph{oblivious transfer} (OT)~\cite{Rabin81}. At a very high level,
and focusing on a single input bit, OT allows $\Ptwo$ to retrieve from
$\Pone$ exactly one value in pair ($w_{21}^0$, $w_{21}^1$) without
$\Pone$ learning which value was retrieved. After running the OT
protocol for every input bit, $\Ptwo$ can evaluate $\mathcal{C}_f$, as
above, to finally compute and send back to $\Pone$ the correct output
$f(x_1,x_2)=1$, deduced by the final garbled value $w_C^1$ of the
output wire.



%
%
%
%
%


\section{Secure $\min$-selection protocols}\label{app:sub}

Here, we overview the design of GC-based protocols $\argMinSelect$ and $\maxDist$/$\minDist$ for secure selection of min/max values, or their index/location, over secret-shared data. These protocols have been defined in Sections~\ref{sec:main} 
and~\ref{sec:analysis} and comprise integral components of our solutions. We provide the exact two circuits over which we can directly apply the garbled-circuits framework (see Appendix~\ref{app:gc}) to get GC-protocols $\argMinSelect$ and $\minDist$, noting that the circuit in support of $\maxDist$ is similar to the case of $\minDist$.

Recall that data consists of $\lambda$-bit values and is secret shared among the two parties as $\kappa$-bit random blinding terms, $\kappa>\lambda$, and $\kappa+1$-bit blinded values, each resulted by adding a random blinding term to an ordinary data value.

Our circuits use as building blocks the following gates, efficient implementations of which are well studied~\cite{min1}:

\begin{itemize}
\item \textsf{ADD}/\textsf{SUB} that adds/subtracts $\kappa+1$-bit integers;

\item \textsf{MIN}/\textsf{MAX} that selects the $\min/\max$ of two $\lambda$-bit integers, using a one-bit output to encode which input value is the $\min/\max$ value (e.g., on input $3,5$ \textsf{MIN} outputs 0 to indicate the first value is smaller); 

\item a multiplexer gate \textsf{MUX}$_i$ that on input two $i$-bit inputs and a selector bit $s$, outputs the first or the second one, depending on the value of~$s$; and finally

\item hard-coded in the circuit constant gates \textsf{CON}$_i$, $1\leq i \leq n$, that always output the $(\log n)$-bit fixed value $i$ (e.g., \textsf{CON}$_3$ outputs the binary representation of~$3$).
\end{itemize}

\begin{figure}[tbph!]
	\centering
	\includegraphics[width=.7\linewidth]{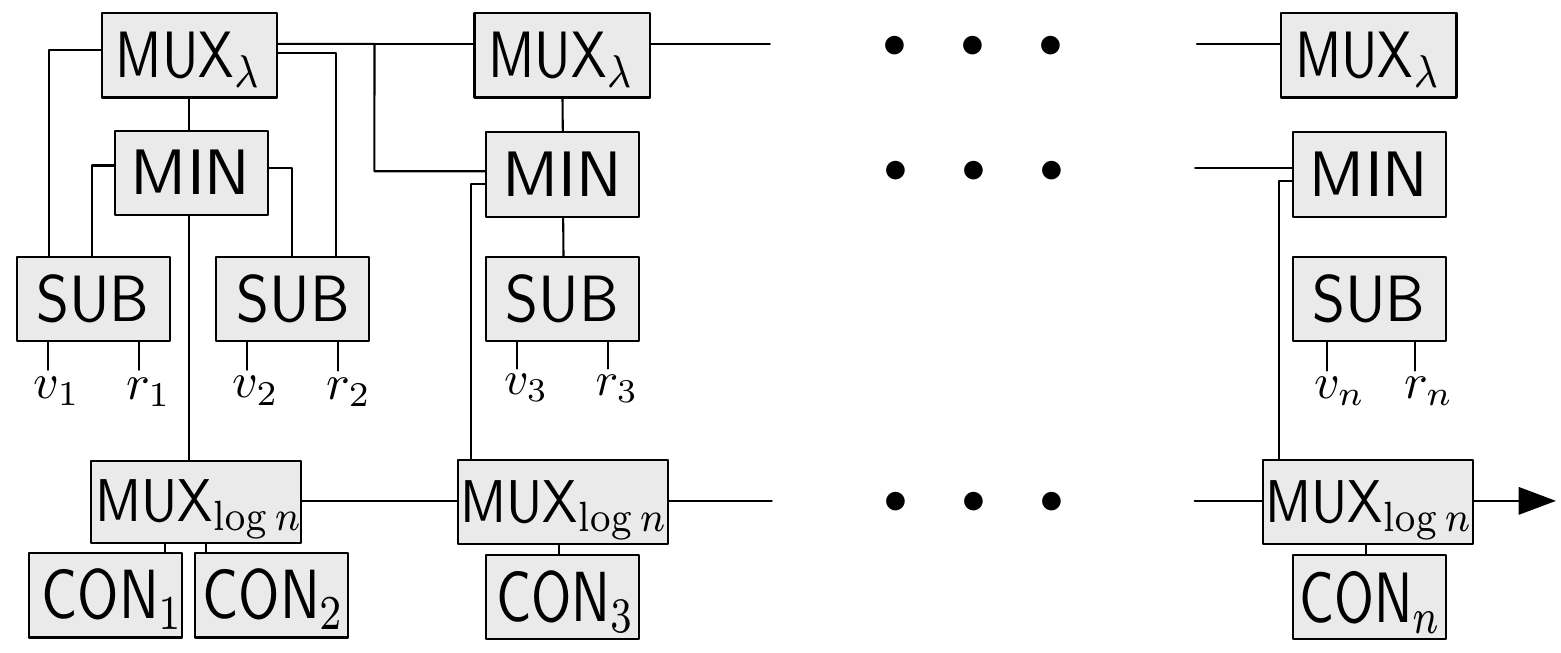}
	\caption{The circuit for protocol $\mathsf{\argMinSelect}$.}\label{fig:sub2}
\end{figure}

Figure~\ref{fig:sub2} shows the circuit of protocol $\mathsf{\argMinSelect}$ for selecting the index of the minimum value in an array of $n$ different values. On input $n$ $\kappa+1$-bit values $v_1,\dots,v_n$ and $n$ $\kappa$-bit blinding terms $r_1,\dots,r_n$, the circuit first uses $n$ \textsf{SUB} gates to compute (the secret) values $v_i - r_i$, $i=1,\dots,n$, and then selects the index of the minimum such value in $n-1$ successive comparisons as follows. In the $i$th comparison, a \textsf{MIN} gate compares the currently minimum value $m_i$ of index $loc_i$ (initially, $m_1=v_1-r_1$, $loc_1=1$) to value $v_{i+1}-r_{i+1}$ of index $i+1$, and its output bit is fed, as the selector bit, to two multiplexer gates \textsf{MUX}$_{\mu}$:

\begin{itemize}
\item $\mu=\log n$: once for selecting among two $(\log n)$-bit indices $loc_i$ and $i+1$, the latter conveniently encoded as the output of constant gate \textsf{CON}$_{i+1}$ (such hard-coded indices significantly facilitate their propagation in the circuit, compared to the alternative of handling indexes as input and carrying them over throughout the circuit); and
\item $\mu=\lambda$: once for selecting among two $\lambda$-bit values $m_i$ and $v_{i+1}-r_{i+1}$,
\end{itemize}
overall propagating the updated minimum value $m_{i+1}=\min\{a, b\}$ and its index $loc_{i+1}$ to the next $(i+1)$th comparison.  The final output (see arrow wire) corresponds to the output of the $(n-1)$th index-selection multiplexer gate.

\begin{figure}[tbph!]
	\centering
	\includegraphics[width=.25\linewidth]{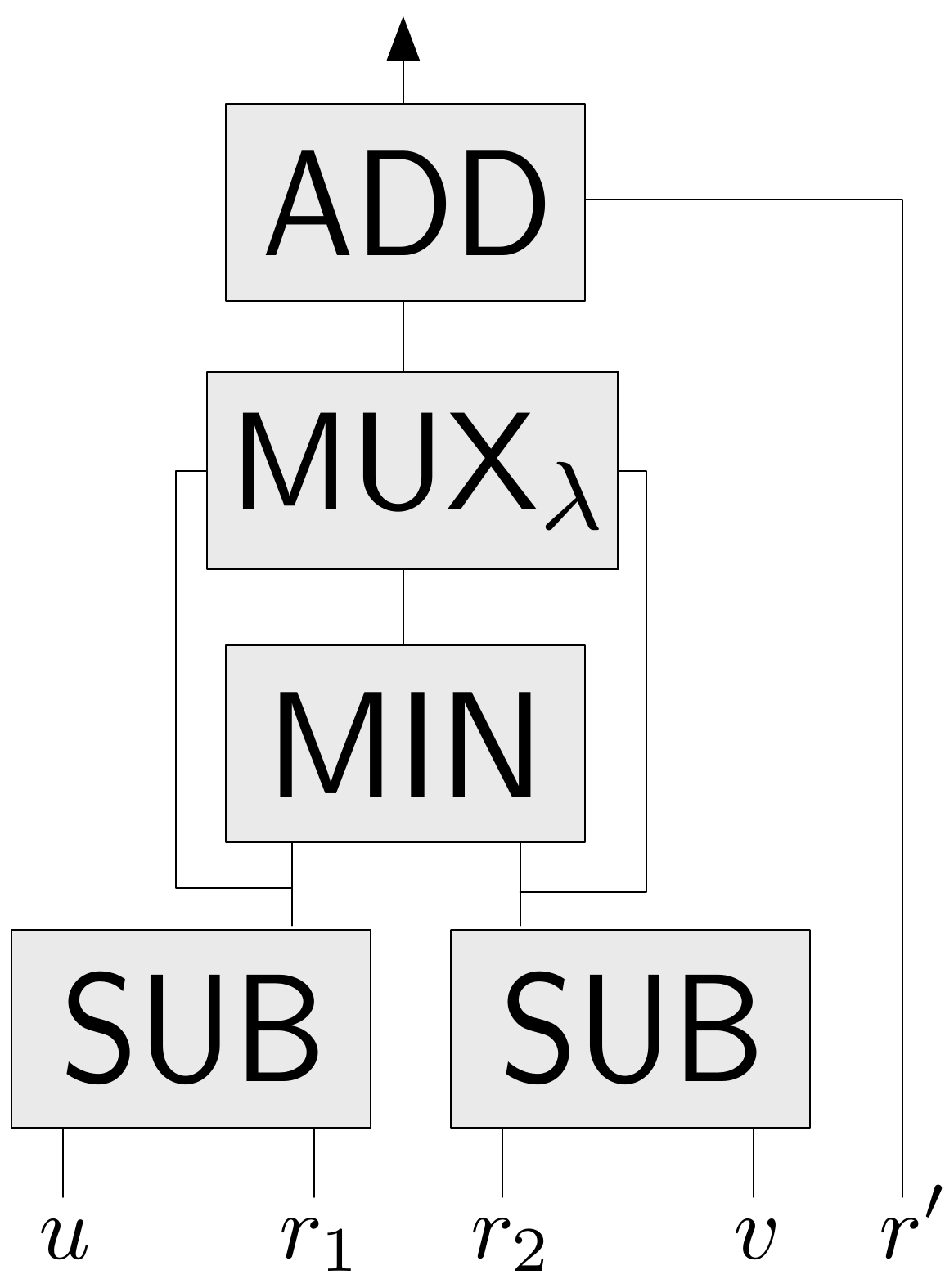}
	\caption{The circuit for protocol $\minDist$.}\label{fig:sub1}
\end{figure}

Figure~\ref{fig:sub1} shows the circuit for protocol $\minDist$ for selecting and re-blinding the minimum value among two secret-shared values. On input two $\kappa+1$-bit blinded values $u,v$ and three $\kappa$-bit blinding terms $r_1,r_2,r'$, the circuit first computes $u-r_1$, $v-r_2$ using two \textsf{SUB} gates, then computes the minimum of these two values using a \textsf{MIN} gate, and its output bit is fed, as the selector bit, to a multiplexer gate \textsf{MUX}$_{\lambda}$ for selecting the minimum among two $\lambda$-bit values $u-r_1$ and $v-r_2$, which is becomes the final output (see arrow wire) after it is blinded by adding the input blinding term $r'$ through a \textsf{ADD} gate. (The circuit for protocol $\maxDist$ is the same with a \textsf{MAX} gate replacing the \textsf{MIN} gate.)


\section{Proof of Theorem~\ref{THM:MAIN}}\label{app:proofs}
We begin by recalling that, under the assumption that the oblivious transfer protocol used is secure, there exists simulator $\simu_{OT}$ that can simulate the views of each of the parties $\Pone,\Ptwo$ during a single oblivious transfer execution when given as input the corresponding party's input (and output, in case it is non-empty) and randomness.

The core idea behind our proof is that, since all values seen by the two parties during the protocol execution (apart from the indexes of the merged clusters at each clustering round) are ``blinded'' by large random factors, these values can be perfectly simulated, as needed in our proof, by randomly selected values. For example, assuming all values $p_i,q_i$ are 32-bits and the chosen random values are 100-bits, it follows that the sum of the two is statistically indistinguishable from a 100-bit value chosen uniformly at random. In particular, this allows the simulator to effectively run the protocol with the adversary by simply choosing simulated values for the other party which he chooses himsellf at random (in the above example these would be random 32-bit values).

We handle the two cases of party corruption separately.

\smallskip
\noindent\textbf{Corruption of $\Ptwo$.} The view of $\Ptwo$ during the protocol execution consists of:
\begin{enumerate}
	\item Encrypted matrices $B,R$ and encrypted arrays $L,S$.
	\item For each clustering round $\ell$, messages received during the oblivious transfer execution for $\mathsf{ArgMin}$, denoted by $OT_{\ell}$ and the min/max index $\alpha_{\ell}$.
	\item During each clustering round $\ell$, for each execution of $\minDist$/$\maxDist$ for index $k$, messages received during the corresponding oblivious transfer execution, denoted by $OT_{\ell,k}$, corresponding garbled circuit $GC_{\ell,k}$, and output value $v_{\ell,k}$.
	\item Encrypted cluster representative values $E_1,\dots,E_{\ell_t}$.
\end{enumerate}

The simulator $\simu_{P_2}$, on input the random tape $R_2$, points $q_1,\dots,q_{n_2}$, outputs $(rep_1/|J_1|,|J_1|,\dots,rep_{\ell_t}/|J_{\ell_t}|,|J_{\ell_t}|)$, $\alpha_1,\dots,\alpha_{\ell_t}$, computes  the view of $\Ptwo$ as follows.

\begin{itemize}
	\item \textbf{(Ciphertext computation)}
          Using random tape $R_2$, the simulator runs the key generation algorithm for $\Ptwo$ to receive $sk',pk'$. He then chooses values $p'_{1},\dots,p'_{n_1}$ uniformly at random from $\{0,1\}^d$. These will act as the ``simulated'' values for player $\Pone$. He then runs protocol $\hcluster$ honestly using the values $p'_i$ as input for $\Pone$ (and the actual values $q_i$ of $\Ptwo$), with the following modifications.
	\item \textbf{(Oblivious transfer simulation for $OT_{\ell}$)} For $\ell=1,\dots,\ell_t$ let $W_\ell$ be the set of garbled input values computed by $\Ptwo$ for the garbled circuit that evaluates $\minDist$/$\maxDist$ at clustering round $\ell$. Since we are in the semi-honest setting, the corrupted $\Ptwo$ computes these values uniformly at random. Therefore, the simulator can also compute them using $R_2$. Then, for $i=1,\dots,\ell,$ the simulator includes in the view (instead of $OT_\ell$) the output $OT'_\ell$ produced by simulator $\simu_{OT}^{(2)}$ on input $W_\ell$.\footnote{\scriptsize And corresponding randomness derived from $R_2$.} Note that $\Ptwo$ does not receive any output from this oblivious transfer execution, thus $\simu_{OT}^{(2)}$ only works given the input.

	\item\textbf{(Oblivious transfer simulation for $\mathsf{Argmin}$)} For each clustering round $\ell$, the simulator includes in the view, the index $\alpha_\ell$.
	
	\item \textbf{(Garbled circuit simulation for $GC_{\ell,k}$)} Next, the simulator needs to compute the garbled circuits $GC_{\ell,k}$. The simulator uses the corresponding values from $\R$ (as computed so far) and a  ``new'' blinding factor $\rho_{\ell,k}$ for $\Pone$' inputs and computes a garbled circuit for evaluating $\mathsf{ArgMin}$ honestly.
	The simulator also includes in the view of $\Ptwo$ the garbled inputs for the corresponding elements from $\R$.
		
	\item \textbf{(Oblivious transfer simulation for $OT_{\ell,k}$)} Let $y_{\ell,k}$ be the input of $\Ptwo$ for the circuit $GC_{\ell,k}$ (i.e., the execution of $\argMinSelect$ for index $k$ during clustering round $\ell$). Since we are in the semi-honest case, the corrupted $\Ptwo$ will provide as input the values that have been established from the interaction with $\Pone$ (using the points $p'_i$) up to that point, therefore $y_{\ell,k}$ can be computed by the simulator. In order to compute the parts of the view that correspond to each of $OT_{\ell,k}$ the simulator includes in the view the output of $\simu_{OT}$ on input $y_{\ell,k}$ and the corresponding choice from each pair of garbled inputs he chose in the previous step (as dictated by the bit representation of $y_{\ell,k}$), which we denote as $OT'_{\ell,k}$.
	
    \item \textbf{(Encrypted representatives computation)} For $\ell=1,\dots,\ell_t$, the simulator computes $rep_\ell = \lceil rep_\ell/|J_\ell| \cdot |J_i| \rceil$ and $E_\ell = \en{rep_\ell}$, where encryption is under (the previously computed) $pk$.
\end{itemize}

We now argue that the view produced by our simulator is indistinguishable from the view of $\Ptwo$ when interacting with $\Pone$ running  $\hcluster$. This is done via the following sequence of hybrids.

\noindent\textbf{Hybrid 0.} This is the view $\view_{\adv^{\hcluster}_{P_2}}$, i.e., the view of $\Ptwo$ when interacting with $\Pone$ running  $\hcluster$  for points $p_i$.

\noindent\textbf{Hybrid 1.} This is the same as Hybrid 0, but the output of $GC_{\ell}$ in $\view_{\adv^{\hcluster}_{P_2}}$ is replaced by $\alpha_{\ell}$. This is indistinguishable from Hybrid 0 due to the correctness of the garbling scheme. Since we are in the semi-honest setting, both parties follow the protocol, therefore the outputs they evaluate are always~$\alpha_{\ell}$.

\noindent\textbf{Hybrid 2.} This is the same as Hybrid 1,  but values in $B,L$ are computed using values $p'_i$. This is statistically indistinguishable from Hybrid 1 (i.e., even unbounded algorithms can only distinguish between the two with probability $O(2^{\kappa}$) since in $\view_{\adv^{\hcluster}_{P_2}}$, each of the values in $B,L$ are computed as the sum of a random value from $\{0,1\}^\kappa$ and a distance between two clusters.

\noindent\textbf{Hybrid 3.} This is the same as Hybrid 2, but all values in $R,S$ are replaced with encryptions of zero's. This is indistinguishable from Hybrid 2 due to the semantic security of Paillier's encryption scheme.

\noindent\textbf{Hybrid 4.} This is the same as Hybrid 3, but each of $OT_{\ell}$ is replaced by $OT'_{\ell}$, computed as described above. This is indistinguishable from Hybrid 3 due to the security of the oblivious transfer protocol.

\noindent\textbf{Hybrid 5.} This is the same as Hybrid 4, but the garbled inputs given to $\Ptwo$ for $GC_{\ell,k}$ are chosen based on the values that have been computed using values $p'_i$. Since garbled inputs are chosen uniformly at random (irrespectively of the actual input values), this follows the same distribution as Hybrid 3.

\noindent\textbf{Hybrid 6.} This is the same as Hybrid 5, but each of $OT_{\ell,k}$ is replaced by output of $OT_{\ell,k}$ computed as described above. This is indistinguishable from Hybrid 5 due to the security of the oblivious transfer protocol.

\noindent\textbf{Hybrid 7.} This is the same as Hybrid 6, but each value $E_i$ sens to $\Ptwo$ is computed as $\en{\lceil rep_i/|J_i| \cdot |J_i| \rceil}$ using public key $pk'$. This is indistinguishable from Hybrid 6 since we are in the semi-honest setting and both parties follow the protocol therefore the outputs they evaluate are always $rep_i/|J_i|$.

Note that Hybrid 7 corresponds to the view produced by our simulator and Hybrid 0 to the view that $\Ptwo$ receives while interacting with $\Pone$ during $\pi_{HC}$ which concludes this part of the proof.

\medskip
\noindent\textbf{Corruption of $\Pone$.} The case where $\Pone$ is corrupted is somewhat simpler as he does not receive any outputs from the circuits $GC_{\ell,k}$. The view of $\Pone$ during the protocol execution consists of:
\begin{enumerate}
	\item Encrypted tables $D,R$ and encrypted arrays $H,L,S$.
	\item For each clustering round $\ell$, a garbled circuit $GC_\ell$  for evaluating $\mathsf{ArgMin}$, messages received during the corresponding oblivious transfer execution denoted by $OT_{\ell}$.
	
	\item During each clustering round $\ell$, for each execution of $\minDist$/$\maxDist$ for index $k$, messages received during the corresponding oblivious transfer execution denoted by $OT_{\ell,k}$.
\end{enumerate}

The simulator $\simu_{P_1}$, on input the random tape $R_1$, points $p_1,\dots,p_{n_1}$, outputs $(rep_1/|J_1|,|J_1|,\dots,rep_{\ell_t}/|J_{\ell_t}|,|J_{\ell_t}|)$, $\alpha_1,\dots,\alpha_{\ell_t}$, computes  the view of $\Pone$ as follows.

\begin{itemize}
	\item \textbf{(Ciphertext computation)}
	Using random tape $R_1$, the simulator runs the key generation algorithm for $\Pone$ to receive $sk,pk$ and computes a pair $sk',pk'$ for himself. He computes $D,H,L$ consisting of encryptions of zeros under $pk'$.
	Moreover, he computes $R$,$S$ consisting of encryption of values chosen uniformly at random from $\{0,1\}^\kappa$ and encrypted under $pk$.
	
	\item\textbf{(Garbled circuit simulation for $GC_{\ell}$)} Next the simulator needs to provide garbled circuits for the evaluation of $\mathsf{ArgMin}$ for each clustering round $\ell$. For this, the simulator creates a ``rigged'' garbled circuit $GC'_{\ell}$ that always outputs $\alpha_{\ell}$, irrespectively of the inputs. This is achieved by forcing all intermediate gates to always return the same garbled output and by setting the output translation temple to always to decode to the bit-representation of $\alpha_{\ell}$ (this process is explained formally in~\cite{GC1}). 	
	
   \item\textbf{(Oblivious transfer simulation for $\mathsf{ArgMin}$)} 	Let $W^{(1)}_{\ell}$, $W^{(2)}_{\ell}$ be the sets of pairs of input garbled values that the simulator choses while creating $GC'_{\ell}$ as described above (where the former corresponds to the input of $\Pone$ and the latter to the input of $\Ptwo$). The simulator includes in the view a random choice from each pair in $W^{(2)}$. Moreover, he replaces the messages in the view that correspond to the execution of $OT_{\ell,k}$, by the output of $\simu_{OT}^{(1)}$ on input $(y_{\ell},W^{(1)}_{\ell})$, where $y_\ell$ is the bit description of the input of $\Pone$ for $GC_\ell$ (which can be computed with the simulator since he has access to $p_i$, $R_1$).

	\item \textbf{(Oblivious transfer simulation for $\minDist$/$\maxDist$)} For each $GC_{\ell,k}$ let $W_{\ell,k}$ be the set of garbled input values computed by $\Pone$ for the garbled circuit that evaluates $\minDist$/$\maxDist$ at clustering round $\ell$ and cluster $k$. Since we are in the semi-honest setting, the corrupted $\Pone$ computes these values uniformly at random. Therefore, the simulator can also compute them using random tape $R_1$. Then, for each $\ell,k$ the simulator includes in the view (instead of $OT_{\ell,k}$) the output $OT'_{\ell,k}$ produced by simulator $\simu_{OT}^{(1)}$ on input $W_{\ell,k}$ (and corresponding randomness derived from $R_1$). Note that $\Pone$ does not receive any output from this oblivious transfer execution, thus $\simu_{OT}^{(1)}$ only works given the input.
	
\end{itemize}

We now argue that the view produced by our simulator is indistinguishable from the view of $\Pone$ when interacting with $\Ptwo$ running $\hcluster$. This is done via the following sequence of hybrids.

\noindent\textbf{Hybrid 0.} This is the view $\view_{\adv^{\hcluster}_{P_1}}$, i.e., the view of $\Pone$ when interacting with $\Ptwo$ running $\pi_{HC}$ for points $q_i$.

\noindent\textbf{Hybrid 1.} This is the same as Hybrid 0, but all values in $D,H',L$ are replaced with encryptions of zero's. This is indistinguishable from Hybrid 1 due to the semantic security of Paillier's encryption scheme.

\noindent\textbf{Hybrid 2.} This is the same as Hybrid 1,  but values in $R,S$ are computed as encryptions of values chosen uniformly at random from $\{0,1\}^\kappa$ under key $pk$. This is statistically indistinguishable from Hybrid 1  for the same reasons as for the case of $\Ptwo$ above.

\noindent\textbf{Hybrid 3.} This is the same as Hybrid 2, but each of $GC_{\ell}$ is replaced by $GC'_{\ell}$, computed as described above (including the values from $W_{(2)})$ This is indistinguishable from Hybrid 2 due to the security of encryption scheme used for the garbling scheme (this is formally described in~\cite{GC1}).

\noindent\textbf{Hybrid 4.} This is the same as Hybrid 3, but each of $OT_{\ell}$ is replaced by $OT'_{\ell}$, computed as described above. This is indistinguishable from Hybrid 3 due to the security of the oblivious transfer protocol.

\noindent\textbf{Hybrid 5.} This is the same as Hybrid 4, but each of $OT_{\ell,k}$ is replaced by $OT'_{\ell,k}$ computed as described above. This is again indistinguishable from Hybrid 5 due to the security of the oblivious transfer protocol.


Note that Hybrid 5 corresponds to the view produced by our simulator and Hybrid 0 to the view that $\Ptwo$ receives while interacting with $\Pone$ during $\hcluster$ which concludes this part of the proof.




\end{document}